\documentclass{emulateapj}
\usepackage{apjfonts}

\usepackage{natbib}
\usepackage{epsfig}

\newcommand{\be}{\begin{equation}}
\newcommand{\ee}{\end{equation}}

\newcommand{\eg}{\emph{e.g.}}
\newcommand{\kms}{\mbox{km\ \ensuremath{\rm{s}^{-1}}}}
\newcommand{\baboon}{IRAS~19312+1950}
 
\shortauthors{Cordiner et al.}

\begin{document}

\title{On the nature of the enigmatic object IRAS~19312+1950:\\ A rare phase of massive star formation?}

\author{M. A. Cordiner\altaffilmark{1}$^,$\,\altaffilmark{2}, A. C. A. Boogert\altaffilmark{3}, S. B. Charnley\altaffilmark{1}, K. Justtanont\altaffilmark{4}, N. L. J. Cox\altaffilmark{5}$^,$\,\altaffilmark{6},\\ R. G. Smith\altaffilmark{7}, A. G. G. M. Tielens\altaffilmark{8}, E. S. Wirstr{\"o}m\altaffilmark{4}, S. N. Milam\altaffilmark{1} and J. V. Keane\altaffilmark{9}}

\altaffiltext{1}{Astrochemistry Laboratory, NASA Goddard Space Flight Center, Code 691,
8800 Greenbelt Road, Greenbelt, MD 20771, USA.}
\email{email:martin.cordiner@nasa.gov}
\altaffiltext{2}{Department of Physics, The Catholic University of America, Washington, DC 20064, USA.}
\altaffiltext{3}{Universities Space Research Association, Stratospheric Observatory for Infrared Astronomy, NASA Ames Research Center, MS 232-11, Moffett Field, CA 94035, USA.}
\altaffiltext{4}{Department of Earth and Space Sciences, Chalmers University of Technology, Onsala Space Observatory, SE-439 92, Onsala, Sweden.}
\altaffiltext{5}{Instituut voor Sterrenkunde, KU~Leuven, Celestijnenlaan 200D, bus 2401, B-3001, Leuven, Belgium}
\altaffiltext{6}{Current address: Universit{\'e} de Toulouse, UPS-OMP, IRAP, F-31028 Toulouse, France.}
\altaffiltext{7}{School of Physical, Environmental \& Mathematical Sciences, The University of New South Wales, Australian Defence Force Academy, Canberra ACT 2600, Australia.}
\altaffiltext{8}{Leiden Observatory, University of Leiden, P.O. Box 9513, NL-2300 RA Leiden, The Netherlands.}
\altaffiltext{9}{Institute for Astronomy, University of Hawaii, Honolulu, HI 96822, USA.\\[2mm]{\it Herschel} is an ESA space observatory with science instruments provided by European-led Principal Investigator consortia and with important participation from NASA.}

\begin{abstract}

IRAS~19312+1950 is a peculiar object that has eluded firm characterization since its discovery, with combined maser properties similar to an evolved star and a young stellar object (YSO). To help determine its true nature, we obtained infrared spectra of IRAS~19312+1950 in the range 5-550~$\mu$m using the {\it Herschel} and {\it Spitzer} space observatories. The {\it Herschel} PACS maps exhibit a compact, slightly asymmetric continuum source at 170~$\mu$m, indicative of a large, dusty circumstellar envelope. The far-IR CO emission line spectrum reveals two gas temperature components: $\approx0.22M_{\odot}$ of material at $280\pm18$~K, and $\approx1.6M_{\odot}$ of material at $157\pm3$~K. The O\,{\sc i} 63~$\mu$m line is detected on-source but no significant emission from atomic ions was found. The HIFI observations display shocked, high-velocity gas with outflow speeds up to 90~\kms\ along the line of sight. From {\it Spitzer} spectroscopy, we identify ice absorption bands due to H$_2$O at 5.8~$\mu$m and CO$_2$ at 15~$\mu$m. The spectral energy distribution is consistent with a massive, luminous ($\sim2\times10^4L_{\odot}$) central source surrounded by a dense, warm circumstellar disk and envelope of total mass $\sim500$-$700M_{\odot}$, with large bipolar outflow cavities. The combination of distinctive far-IR spectral features suggest that IRAS~19312+1950 should be classified as an accreting high-mass YSO rather than an evolved star. In light of this reclassification, IRAS~19312+1950 becomes only the 5th high-mass protostar known to exhibit SiO maser activity, and demonstrates that 18 cm OH maser line ratios may not be reliable observational discriminators between evolved stars and YSOs.

\end{abstract}


\section{Introduction}

\baboon\ is an infrared-bright object located in the Galactic plane at a distance of about $3.8$~kpc \citep{ima11}. Observations by 2MASS show an extended, horn-like IR nebulosity surrounding the central (point-like) source \citep{nak00}. A more recent near-IR (JHK) image from the UKIRT Infrared Deep Sky Survey is shown in Figure \ref{fig:ukidss}, in which a complex filamentary structure is evident, spanning a scale of $\sim30''$.

Based on SiO and H$_2$O maser detections, \citet{nak00} deduced that the central source of \baboon\ is most likely an oxygen-rich evolved star similar to OH\,231.8+4.2. Subsequent molecular line observations by \citet{deg04} identified abundant carbon, nitrogen and oxygen-bearing molecules in the circumstellar envelope, many of which show a complex kinematical structure. The presence of a strong, narrow, bipolar emission component (about 1-2~km\,s$^{-1}$ wide), superimposed on a broader component with FWHM $\sim50$~km\,s$^{-1}$ led \citet{deg04} and \citet{nak05} to interpret the observed emission as arising in the outflows from an AGB stellar atmosphere.  However, several observational characteristics of \baboon\ are unusual for an AGB star, which justifies a closer look at the nature of this object: (1) the linear distribution of H$_2$O maser spots and complex OH maser line profile \citep{nak11}; (2) the detection of common interstellar molecules, including CH$_3$OH, N$_2$H$^+$ and HC$_3$N, which are not normally seen in O-rich AGB star envelopes; and (3) the association of \baboon\ with a massive, dense molecular cloud core G055.372+00.185 \citep{dun11}.

\citet{nak11} identified two most likely scenarios to explain the observations of \baboon. First, they considered the central object to be a massive ($\sim10\,M_{\odot}$) O-rich AGB star with a fast bipolar outflow embedded in a chemically-rich molecular cloud. \citet{deg04} noted that the association with a compact molecular cloud is highly unusual, but could be the result of a chance encounter \citep[\eg][]{kas94}. On the other hand, the evidence could be indicative that \baboon\ is an unusual, massive young stellar object, which would be consistent with some of the observed outflow characteristics as well as the high CH$_3$OH and HC$_3$N abundances. A Class~I CH$_3$OH maser was recently detected in \baboon\ by \citet{nak15}. This class of maser has not previously been seen in evolved stars, but is common in regions of high mass star formation as a tracer of molecular outflows \citep[\emph{e.g.}][]{cyg09}. Strongly variable red and blue-shifted 43~GHz SiO maser peaks were observed towards \baboon\ by \citet{sco02} and \citet{nak11}, with a velocity separation of about 30-36~\kms. \citet{sco02} speculated that these masers could arise in the approaching and receding sides of a rotating disk about a young stellar object, similar to Orion KL. SiO masers, however, are extremely rare in young stellar objects, having been seen in only 4 massive star forming regions to-date: Orion KL, Sgr B2, W51 \citep{zap09} and recently in Galactic Center Cloud C, G0.38+0.04 \citep{gin15}. By contrast, SiO masers have been observed in thousands AGB stars. Furthermore, the strong 1612~MHz OH maser line of \baboon\ (relative to the 1665 and 1667~MHz lines) is a typical characteristic of evolved stars \citep{cas98,nak11}, so if confirmed as a YSO, the combined maser properties of \baboon\ would be remarkable. 

The spectral energy distribution (SED) of \baboon\ was modeled by \citet{mur07} assuming the central object to be a high mass-loss AGB star.  At the time of that study, the SED was not well constrained in the far-IR. A severe mis-match between model and observations at the longest wavelengths ($>20~\mu$m) was dismissed as primarily the result of background flux contamination. However, the large far-infrared fluxes measured by IRAS indicate that the circumstellar envelope mass (as well as the intrinsic source luminosity and temperature) may have been underestimated. If confirmed, an SED that rises to a peak in the far-IR would be more reminiscent of a young, massive YSO embedded in a dense, accreting envelope. 

To help determine the true nature of IRAS~19312+1950, we have obtained mid and far-IR spectroscopic observations to probe the properties of the environment surrounding this enigmatic, dust-enshrouded object. The {\it Herschel} Space Observatory \citep{pil10} HIFI (Heterodyne Instrument for the Far-Infrared; \citealt{deg10}) and PACS (Photodetector Array Camera \& Spectrometer; \citealt{pog10}) instruments were used to observe emission from dust, carbon monoxide, water and other species in the wavelength range 51-550~$\mu$m. {\it Spitzer} Space Telescope spectra were also obtained, covering the range 5-35~$\mu$m. The combined {\it Herschel} and {\it Spitzer} observations provide crucial new information on the SED and ice properties, complementing the previous IR data from ISO, 2MASS, Akari, WISE and other telescopes. In addition, our high signal-to-noise, spectrally-resolved  HIFI CO and H$_2$O line profiles provide new information on the nature of the outflow. The evidence provided by these observations strongly favor the identification of IRAS~19312+1950 as an emdedded, high-mass YSO.

\section{Observations and data reduction}

\subsection{{\it Herschel} observations}

\begin{deluxetable*}{ccccccll}
\centering
\tabletypesize{\small}
\tablecaption{\textit{Herschel} Observations of \baboon \label{tab:obs}}
\tablewidth{0pt}
\tablehead{OBSID&Date&RA&Dec.&Instrument&Mode&Setting&Duration (s)}
\startdata
1342245229&2012-05-01&19:33:24.3&+19:56:55&PACS&Range Scan&B2B&9134\\
1342245230&2012-05-01&19:33:24.3&+19:56:55&PACS&Range Scan&B2A&4616\\
1342245377&2012-05-04&19:33:24.3&+19:56:55&HIFI&WBS DBS&1102+1114&3286\\
1342245378&2012-05-04&19:33:24.3&+19:56:55&HIFI&WBS DBS&1095+1111&2028\\
1342245380&2012-05-04&19:33:24.3&+19:56:55&HIFI&WBS DBS&546+558&758\\
1342252110&2012-10-02&19:33:24.3&+19:56:55&HIFI&WBS DBS&647+659&2448\\
1342252111&2012-10-02&19:33:24.3&+19:56:55&HIFI&WBS DBS&679+691&1548\\
1342255788&2012-11-22&19:33:24.3&+19:56:55&HIFI&WBS DBS&576+588&894\\
1342255789&2012-11-22&19:33:24.3&+19:56:55&HIFI&WBS DBS&608+621&739

\enddata
\tablecomments{HIFI settings indicate the central frequencies (in GHz) of the lower and upper receiver sidebands, respectively (each sideband is 4 GHz wide).}
\end{deluxetable*}

\begin{figure}
\centering
\includegraphics[width=\columnwidth]{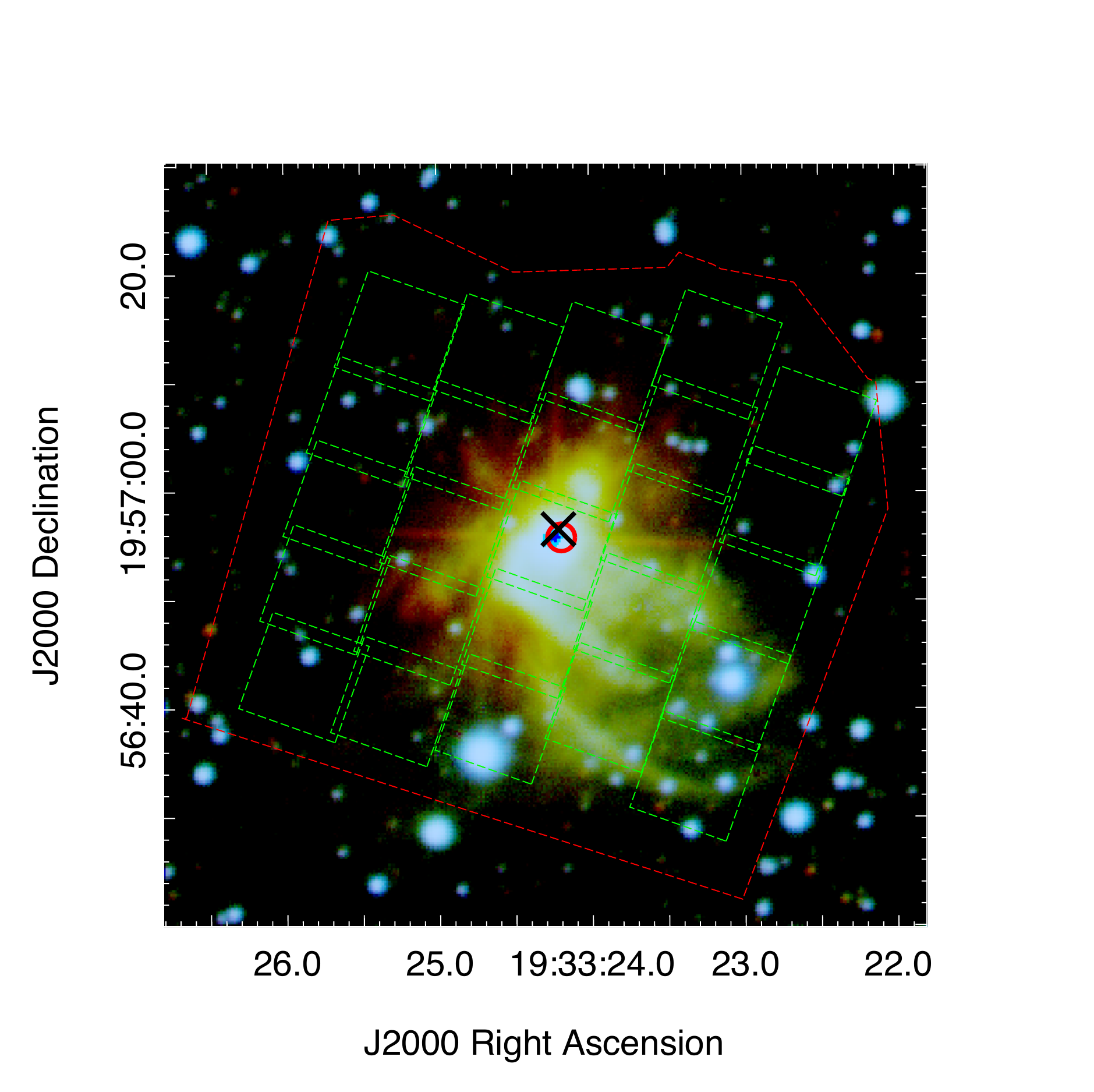}
\caption{UKIDSS $J$ (blue), $H$ (green), $K$ (red) image of IRAS~19312+1950 with $5\times5$ PACS spaxel footprint overlaid (green dashed boxes). Brightness histograms have been stretched to emphasize the IR nebulosity. Each PACS spaxel has dimensions of $9.4''\times9.4''$. The image is centered on the location of the 2MASS point-source. The position of the peak 170~$\mu$\,m continuum flux is shown with a black `$\times$'. Red circle indicates the CO far-IR emission peak. Red dashed outline indicates the ($\approx54''\times54''$) region covered by the PACS raster maps. For details of the UKIDSS project see \citet{law07}. UKIDSS uses the UKIRT Wide Field Camera (WFCAM; \citealt{cas07}); the photometric system and calibration were described by \citet{hew06} and \citet{hod09}, respectively.  \label{fig:ukidss}}
\end{figure}

{\it Herschel} observations of \baboon\ were obtained between May and October 2012 as part of Cycle 2 Open Time Programme \emph{OT2\_mcordine\_2}. Basic observational parameters and instrument settings are given in Table \ref{tab:obs}. Four PACS pointings in a $2\times2$-pointing raster pattern were obtained around the central source position (RA 19:33:24.29, decl. 19:56:55.0 (J2000)), each employing one repetition of the B2A (short) and B2B (long) spectral scan modes to cover the complete wavelength range 51-220~$\mu$m. Observations were obtained in chop/nod mode, using a $6'$ throw either side of the source (at an angle $30^{\circ}$ clockwise from celestial North). The PACS integral-field unit (IFU) consists of a $5\times5$ array of $9.4''$-square spaxels and the spacing between map pointings was $4.5''$ in RA and dec. This ensured full sampling of the spatial point-spread function (PSF), which has a minimum FWHM of about 9$''$ at 50~$\mu$m (rising to about 13$''$ at 180~$\mu$m). The relative locations and shapes of the PACS IFU spaxel array elements are shown overlaid in green over the UKIDSS ($JHK$) image of \baboon\ in Figure \ref{fig:ukidss}.

The PACS range-scan data were reduced using HIPE pipeline version 11.0 \citep{ott10}, which included standard flat-fielding, de-glitching, wavelength and flux calibration routines, as well as projecting of the data into 3-D raster maps with a plate scale of $3''$/pixel. Data at wavelengths $<55$~$\mu$m in the B2A blue spectra, $>98$~$\mu$m in the B2B blue spectra and $>189$~$\mu$m were excluded from subsequent analysis due to contamination by reflected light inside the PACS spectrometer. Absolute Flux calibration is expected to be accurate to within about 10\%.

PACS spectra were extracted by summation over a $54''$-diameter circular aperture at each wavelength. The majority of the flux from \baboon\ (see, \eg, Figure \ref{fig:maps}) is contained in this region.  \baboon\ appears to be the only IR-bright object within the mapped region (also confirmed by 2MASS and UKIDSS imaging), so contamination from other sources is expected to be negligible. To measure the spectral line fluxes, wavelengths were converted to frequencies, then baseline-subtracted using low-order polynomials. Observed line widths of 325~\kms\ at 108~$\mu$m and 212~\kms\ at 174~$\mu$m match the PACS instrument specifications for the spectroscopic resolving power, indicating that the lines were not resolved.

The HIFI instrument was used to obtain single-pointing, dual-polarization, double side-band observations of a selection of transitions from CO and H$_2$O towards the center of \baboon. These were carried out using the Wide-Band Spectrometer (WBS), and the High Resolution Spectrometer (HRS) simultaneously, with spectral resolutions of 1.1~MHz and 0.25~MHz, respectively. Dual beam-switching (DBS) mode was used to remove the instrumental background signal.  The detected lines and their respective wavelengths and frequencies are given in Table \ref{tab:HIFI}.

\begin{deluxetable*}{llcccccrr}
\centering
\tabletypesize{\small}
\tablecaption{HIFI Integrated Line Fluxes \label{tab:HIFI}}
\tablewidth{0pt}
\tablehead{Species&Transition&$\eta_{MB}$&HPBW&$\lambda$&$\nu$&$E_u$&$\int{S_{\nu}d\nu}$&$\int{T_{MB}dv}$\\&&&($''$)&($\mu$m)&(GHz)&(K)&(Jy\,GHz)&(K km\,s$^{-1}$)}
\startdata
CO        &$5-4$& 0.62 &37&520.231&576.267&               82.98 &    79.33 (0.05)  & 126.27 (0.07)\\
CO        &$6-5$& 0.65 &31&433.556&691.473&               116.2 &    115.79 (0.05) & 153.58 (0.07)\\
$^{13}$CO &$6-5$& 0.65 &32&453.498&661.067&               111.1 &    9.16 (0.03)   & 12.72 (0.06) \\
$^{13}$CO &$10-9$& 0.64&19&272.205&1101.35&               290.8 &    4.50 (0.06)   & 3.75 (0.06)  \\
C$^{18}$O  &$6-5$& 0.65&32&455.229&658.553&               110.6 &    1.05 (0.02)   & 1.47 (0.02)  \\
C$^{18}$O  &$10-9$&0.64&19&273.243&1097.16&               289.7 &    0.32 (0.03)   & 0.27 (0.03)  \\
p-H$_2$O  &$1_{11}-0_{00}$&0.64&19&269.272& 1113.34&      53.43 &    21.77 (0.13)  & 17.93 (0.10) \\
o-H$_2$O  &$1_{10}-1_{01}$&0.62&38&538.289& 556.936&      60.97 &    4.96 (0.04)   & 8.18 (0.07)  \\
o-H$_2$O  &$3_{12}-3_{03}$&0.64&19&273.193& 1097.36&      249.4 &    16.03 (0.10)  & 13.39 (0.08) \\
o-H$_2$$^{18}$O&$1_{10}-1_{01}$&0.62&39&547.390& 547.676& 60.46 &    0.20 (0.02)   & 0.33 (0.04)  \\
HCN       &$7-6$&       0.62   &34&483.299&620.304&       119.1 &    1.15 (0.02)   & 1.71 (0.04)  \\
CS        &$12-11$&      0.62   &36&510.184&587.616&      183.4 &    0.09 (0.01)   & 0.13 (0.01)  \\
SiO       &$14-13$&      0.62   &35&493.398&607.608&      218.8 &    0.11 (0.01)   & 0.17 (0.02)
\enddata
\tablecomments{Statistical $\pm1\sigma$ integrated line brightness/flux errors are given in parentheses.}
\end{deluxetable*}

The HIFI spectra were reduced using the HIPE pipeline (version 9.0) then baseline-subtracted using low-order polynomials fitted to the continuum on either side of the lines of interest. Differences in intensity between the two polarizations were negligible, so these were averaged to improve the signal-to-noise ratio.  Spectral line antenna temperatures were corrected to the main beam ($T_{MB}$) scale using the beam efficiencies in Table \ref{tab:HIFI} and a telescope forward efficiency factor of 0.96. Average continuum fluxes from the two sidebands were measured (before baseline subtraction), in regions free from any spectral lines.

\subsection{{\it Spitzer} observations}

Spectra of IRAS 19312+1950 were obtained with the Infrared
Spectrograph (IRS) of the {\it Spitzer} Space
Telescope on 2006 October 19
during IRS campaign 1070 (IRSX007500), as part of
Guaranteed Time Observer program 93 (principal investigator D.
Cruikshank).  The IRS modules Short-Low (SL; 5.2-14.5 $\mu$m at
$R=\lambda/\Delta\lambda\sim $ 100), Long-Low (LL; 14-38 $\mu$m at
$R\sim $ 100) and Short-High (SH; 9.9-19.6 $\mu$m at $R\sim $ 600)
were all observed in staring mode at the default two nodding positions
along the slit and also at two positions on the sky, one centered on
IRAS~19312+1950, and one on a position offset
10$''$ to the East and 10$''$ to the North. In all modules, the
emission in the offset position is negligible compared to the main
target and is therefore not used in the analysis. High pointing accuracy was
guaranteed by using the optical PCRS peak up mode on a nearby
star. Ramp times of 14 and 6 seconds were used for the low (SL and LL)
and high (SH) resolution modules, respectively. SL was observed for 5
cycles, and LL and SH for 3 cycles.

The SL spectra were extracted and calibrated from the two-dimensional
Basic Calibrated Data (BCD) produced by the standard {\it Spitzer}
pipeline (version S18.18.0), using the same method and routines
discussed in \citet{boo11}. Uncertainties (1$\sigma$) for each
spectral point were calculated using the `func' frames provided by
the {\it Spitzer} pipeline. One section of the SL1 sub-module
(11.0-13.0~$\mu$m) suffers from saturation, recognizable by `NaN'
values in the BCD images. The other spectral regions in SL do not
contain saturation flags and are scientifically valid.
All SL data points above 10.0~$\mu$m were removed, however, as this
region is covered by the SH module at higher spectral resolution.  It
was verified that overlapping unsaturated SL and SH data points are in
good agreement. For the SH module, the one-dimensional spectra in the
`tune' tables produced by the {\it Spitzer} pipeline were used. All
data points with non-zero flag values were excluded and the spectra on
the two nodding positions were averaged. This results in a nearly
continuous 9.9-19.6~$\mu$m spectrum not affected by saturation.  The
LL spectra, however, are highly saturated in the peak of the PSF and a special
extraction was performed to mitigate the effects.  Using the {\it Spitzer}
IRS Custom Extraction (SPICE) software package, spectra were extracted
along 3 and 6 pixel-wide rectangles. The resulting spectra were
subtracted from each other, resulting in spectra tracing the regions
1.5-3 pixels (7.7-15.3$''$) on either side of the source. Such
non-standard extraction requires a dedicated spectral response
function. The standard star for the IRS campaign (HR 6606) is weak in
the PSF wings, however, and a new response function could not be
derived. Following the IRS manual, 10\% uncertainties were assigned to
the LL flux values thus derived.

The combined {\it Spitzer} spectrum was produced by multiplying
the SH fluxes by a factor of 0.96 to match SL. The portion of LL
overlapping with SH (14.0-19.6 $\mu$m) was removed, and the longer
wavelengths were multiplied to match SH. Although no emission was detected $10''$ north and east of the source, due to the special extraction method discussed above, it should be kept in mind that the LL spectrum could be contaminated by emission at a distances 7.7-15.3$''$ along the slit.

\section{Results}

\subsection{PACS Range Scans}

\begin{figure*}
\centering
\includegraphics[width=0.45\textwidth]{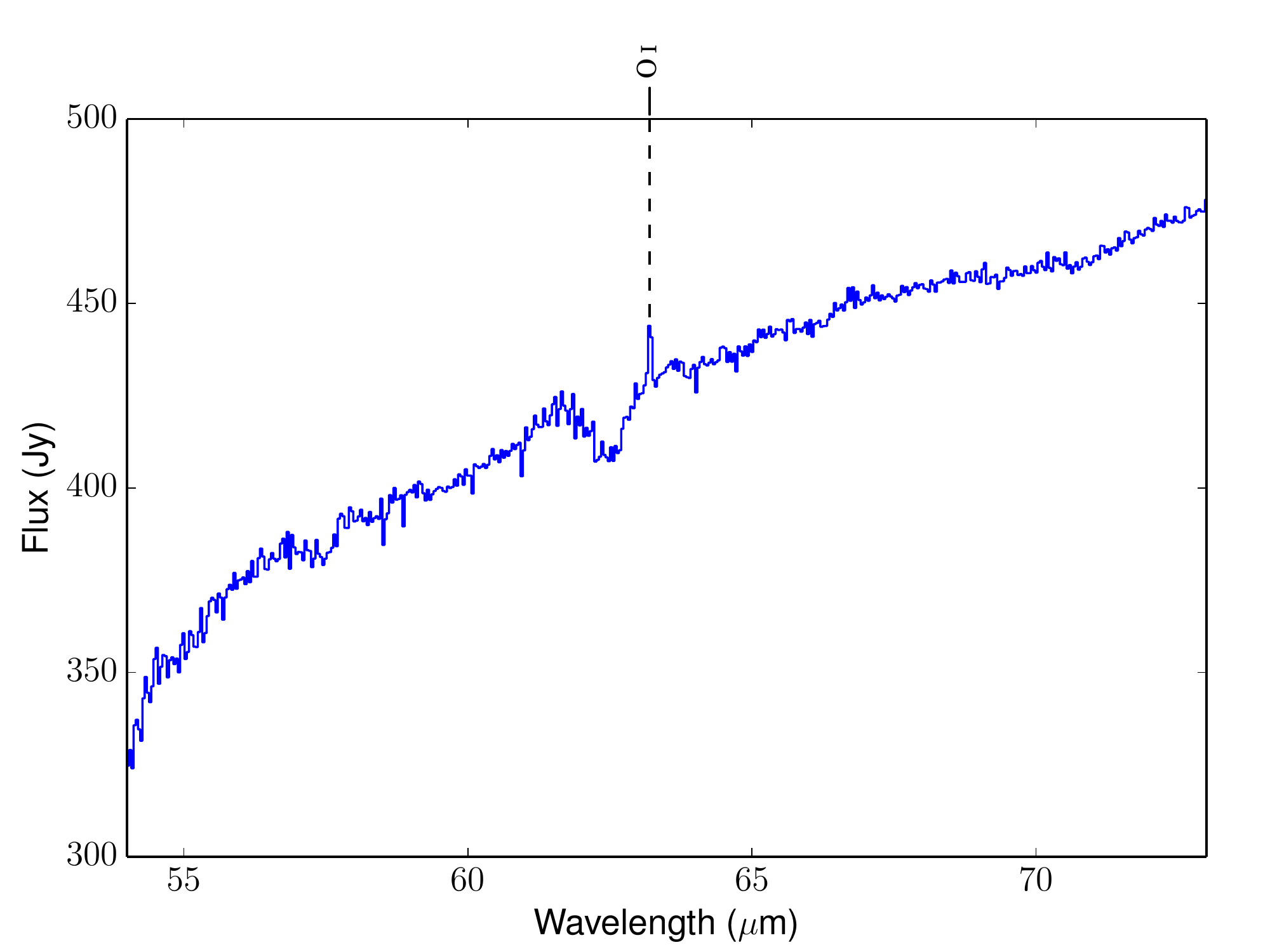}
\includegraphics[width=0.45\textwidth]{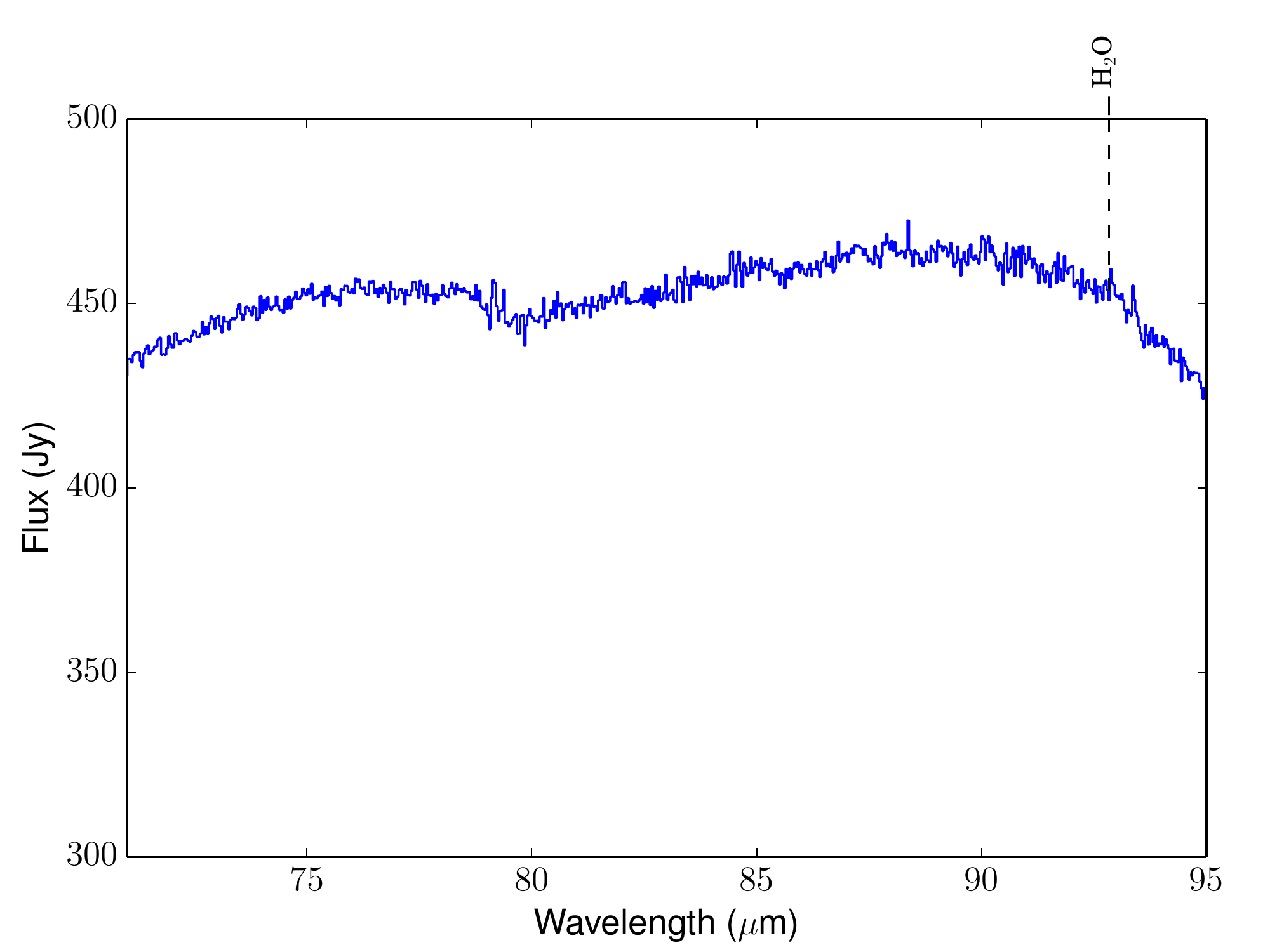}
\includegraphics[width=0.45\textwidth]{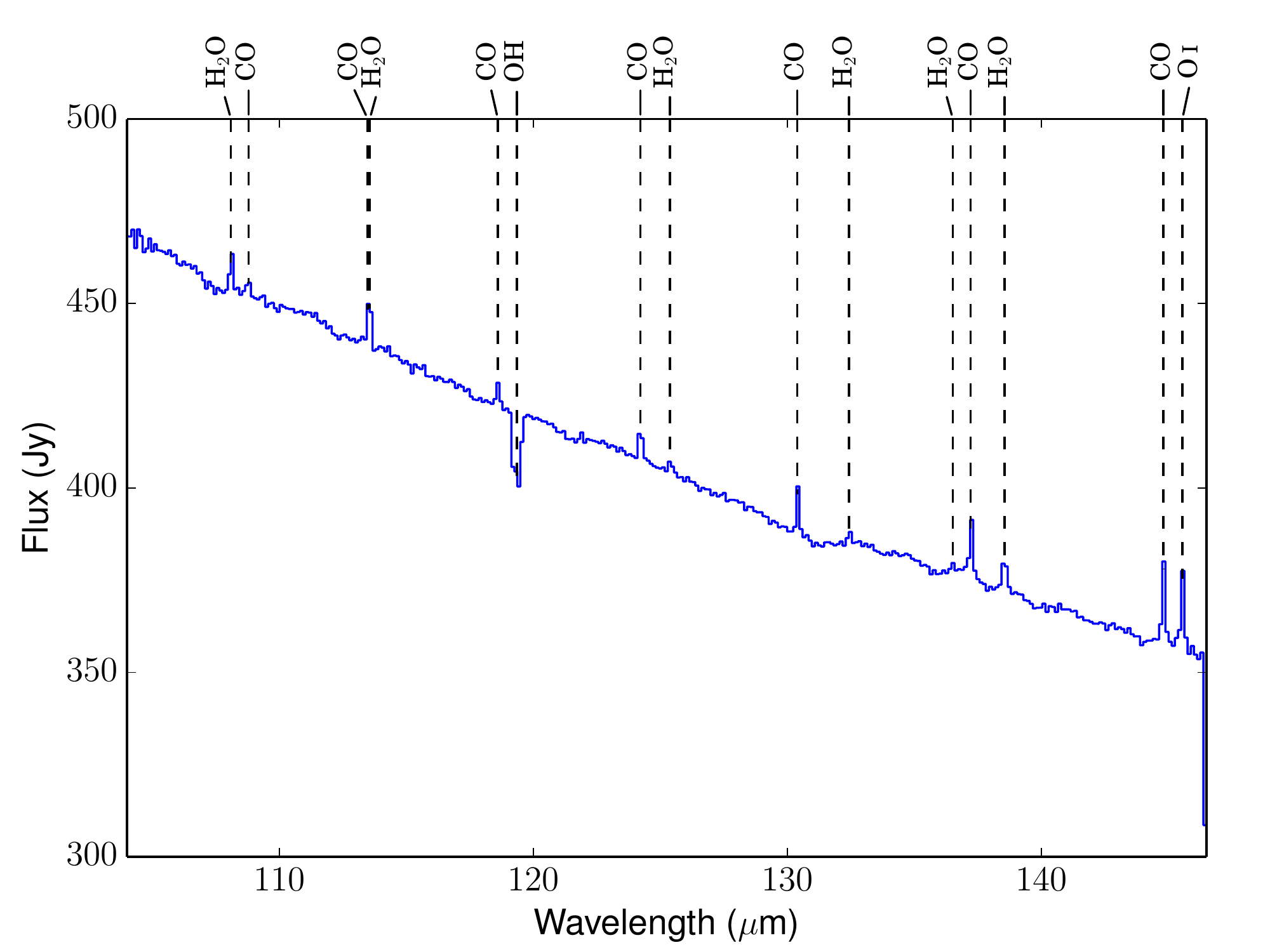}
\includegraphics[width=0.45\textwidth]{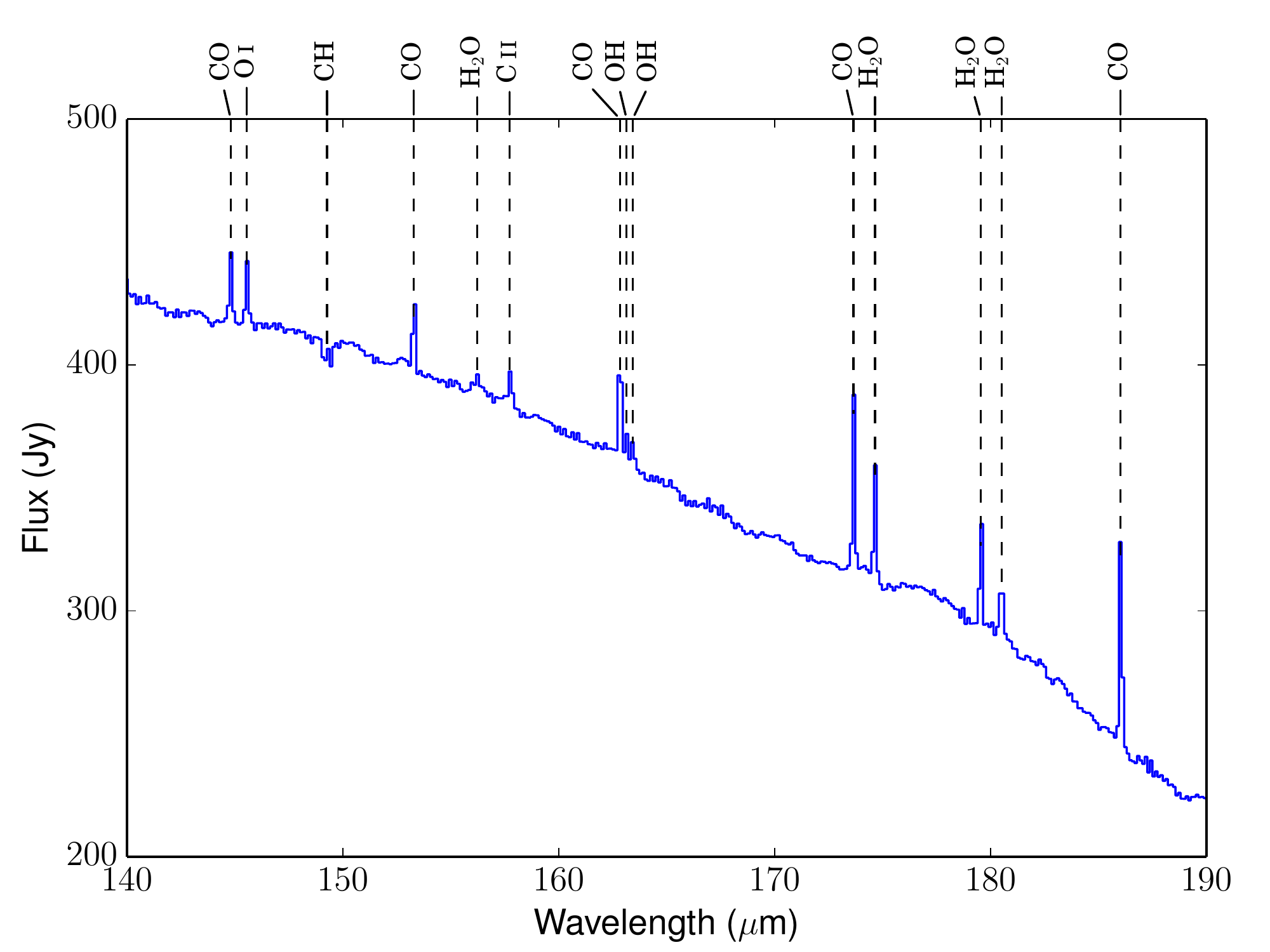}
\caption{PACS range scans integrated over the a $54''$-diameter circular aperture centered on \baboon. Assigned emission and absorption lines are marked. Apparent absorption features around 62.5~$\mu$m and 80~$\mu$m are probably instrumental artifacts.\label{fig:ranges}}
\end{figure*}

PACS spectra in the range 54-189~$\mu$m are shown in Figure \ref{fig:ranges}. A plethora of emission lines is present atop the strong (thermal) dust continuum. Most of the lines are due to H$_2$O and CO; transitions, wavelengths and integrated line fluxes for these species are given in Tables \ref{tab:COfluxes} and \ref{tab:H2Ofluxes}. Integrated line fluxes are in units of Jy\,GHz, with $1\sigma$ statistical errors derived from the RMS noise in the nearby continuum. 

\begin{deluxetable}{lrrr}
\centering
\tabletypesize{\small}
\tablecaption{PACS CO Integrated Line Fluxes\label{tab:COfluxes}}
\tablewidth{0pt}
\tablehead{Transition&$\lambda$ ($\mu$m)&$E_u$(K)&$\int{S_{\nu}d\nu}$ (Jy\,GHz)}
\startdata
14-13&185.999&580.5&133.6 (2.8)\\
15-14&173.631&663.4&112.7 (1.6)\\
16-15&162.812&751.8&89.0 (2.4)\\
17-16&153.266&845.6&63.5 (1.7)\\
18-17&144.784&945.0&78.4 (3.0)\\
19-18&137.196&1050&42.5 (2.3)\\
20-19&130.369&1160&36.6 (1.6)\\
21-20&124.193&1276&30.4 (2.4)\\
22-21&118.581&1397&24.8 (1.9)\\
23-22&113.457&1524&$<69.1$\\
24-23&108.763&1657&15.4 (3.4)
\enddata
\tablecomments{Statistical $1\sigma$ errors are given in parentheses. The $J=23-22$ flux is an upper limit due to blending with the H$_2$O 4$_{14}$-3$_{03}$ transition. 1 Jy\,GHz = $10^{-17}$~W\,m$^{-2}$.}
\end{deluxetable}

\begin{deluxetable}{lrrr}
\centering
\tabletypesize{\small}
\tablecaption{PACS H$_2$O Integrated Line Fluxes \label{tab:H2Ofluxes}}
\tablewidth{0pt}
\tablehead{Transition&$\lambda$ ($\mu$m)&$E_u$(K)&$\int{S_{\nu}d\nu}$ (Jy\,GHz)}
\startdata
6$_{43}$-6$_{34}$&92.811&1089&$<12.5$\\
2$_{21}$-1$_{10}$&108.073&194.1&50.0 (4.8)\\
4$_{14}$-3$_{03}$&113.537&323.5&$<69.1$\\
4$_{04}$-3$_{13}$&125.354&319.5&10.8 (1.8)\\
4$_{23}$-4$_{14}$&132.408&432.2&11.8 (1.8)\\
3$_{30}$-3$_{21}$&136.496&410.7&3.8 (1.8)\\
3$_{13}$-2$_{02}$&138.528&204.7&32.1 (2.3)\\
3$_{22}$-3$_{13}$&156.194&296.8&13.0 (1.8)\\
5$_{32}$-5$_{23}$&160.510&732.1&$<2.8$\\
7$_{34}$-7$_{25}$&166.815&1212&$<2.7$\\
3$_{03}$-2$_{12}$&174.626&196.8&85.0 (2.3)\\
2$_{12}$-1$_{01}$&179.526&114.4&61.0 (2.7)\\
2$_{21}$-2$_{12}$&180.488&194.1&46.7 (3.5)
\enddata
\tablecomments{Statistical $1\sigma$ errors are given in parentheses. Upper limits are $3\sigma\Delta\nu$, except for H$_2$O 4$_{14}$-3$_{03}$, which is blended with CO $23-22$. 1 Jy\,GHz = $10^{-17}$~W\,m$^{-2}$.}
\end{deluxetable}

Emission lines from O\,{\sc i} (at 63~$\mu$m and 146~$\mu$m), C\,{\sc ii} (158~$\mu$m) and an OH doublet (163~$\mu$m) are clearly visible, as well as absorption lines of OH (119~$\mu$m) and CH (149~$\mu$m).  Broad structures in the two shorter-wavelength regions are found to be highly spatially variable across the small region mapped by PACS, and are likely to be a result of instrumental artifacts. All of the lines visible in Figure \ref{fig:ranges} have been identified. We searched for additional lines from species including SiO, HCN, NH$_3$ (and CH$_3$OH), which have previously been seen in the envelopes of oxygen-rich (and carbon-rich) AGB stars using PACS \citep{roy10,dec10}, but no emission from these species was detected. No evidence was found for N\,{\sc ii} or O\,{\sc iii} emission, which are common tracers of ultra-compact H\,{\sc ii} / massive star-forming regions \citep[\eg][]{van10}. The 69~$\mu$m forsterite band (seen in evolved stellar envelopes), was also not detected.

\subsection{CO and H$_2$O rotational diagrams}

\begin{figure}
\centering
\includegraphics[width=\columnwidth]{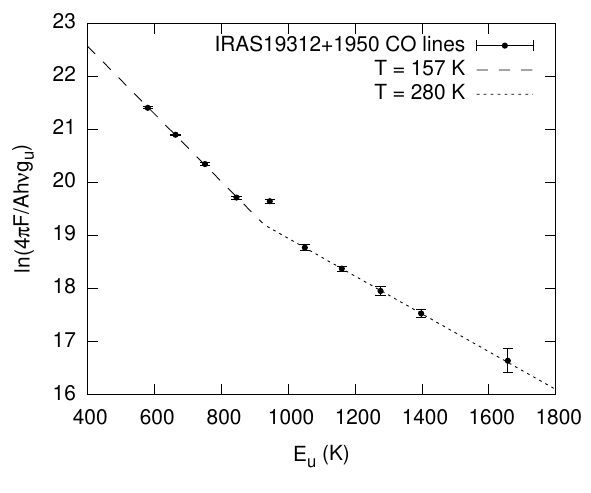}
\includegraphics[width=\columnwidth]{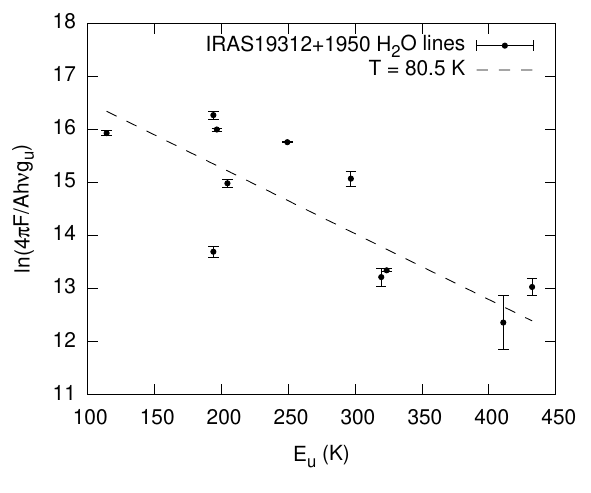}
\caption{Rotational excitation diagram for CO (top) and H$_2$O (bottom) derived from PACS range-scan integrated line fluxes.  \label{fig:rot}}
\end{figure}

Following the formalism described by \citet{jus00}, rotational diagrams were generated for the CO and H$_2$O lines observed with PACS. Neglecting optical depth and background radiation, the spectrally-integrated line flux ($F$) for a transition at frequency $\nu$ is related to the number of molecules in the upper energy state ($N_u$) and the source distance ($d$) by

\be
F=\int{S_{\nu}}d\nu=N_u\frac{Ah\nu}{4\pi d^2}
\ee

\noindent where $A$ is the Einstein spontaneous decay rate. Therefore, in LTE at a temperature $T$,

\be
F=N\frac{e^{-E_u/kT}}{Q(T)}\frac{Ah\nu g_u}{4\pi d^2}
\ee

\noindent where $N$ is the total number of molecules, $E_u$ is the upper-state energy, $g_u$ the degeneracy, and $Q(T)$ is the partition function.  Figure \ref{fig:rot} shows the plots of $\ln(4\pi F/Ah\nu g_u)$ \emph{vs.} $E_u$ for CO and H$_2$O. The gradients of these diagrams are equal to $-1/T$ and the intercepts give $\ln(N/Q(T)d^2)$. Adopting a distance of 3.8~kpc, linear least-squares fits (allowing for errors) result in the following parameters: $T=197.7\pm2.1$~K, $N=(3.3\pm0.2)\times10^{52}$ for CO, and $T=80.5\pm1.4$~K, $N=(1.9\pm0.2)\times10^{49}$ for H$_2$O.  An equilibrium H$_2$O ortho-to-para ratio of 3 has been assumed.

From Figure \ref{fig:rot}, the CO rotational diagram is clearly comprised of two distinct temperature regimes, which indicates the presence of two gas components at different temperatures.  These components were fitted separately, excluding the outlier point at E$_u=945$~K that could be due to non-thermal excitation or blending with an unidentified emission line. The higher-temperature component has $T=280\pm18$~K, $N=8.4^{+3.3}_{-2.3}\times10^{51}$ and the lower-temperature component has $T=157\pm3$~K, $N=6.3^{+0.7}_{0.6}\times10^{52}$.  Assuming a CO/H$_2$ ratio of $10^{-4}$ and a mean mass per particle of 2.4~$amu$, the mass associated with the warmer component is $0.22M_{\odot}$, and with the cooler component $1.6M_{\odot}$.

Rotational diagrams for evolved stellar atmospheres typically exhibit a relatively uniform slope, indicative of a single gas temperature component \citep[\eg][]{jus00,dan14,kho14}. This is consistent with a single dominant heating source in the circumstellar envelope due to infrared irradiation by the dust-enshrouded central star. By contrast, young stellar objects tend to have a more complex heating environment, often resulting in multiple components with different temperatures \citep[\eg][]{fue05,bru09,man13,kar14,lee15}. Three main mechanisms for the excitation of protostellar CO and H$_2$O were considered by \citet{vank10} and \citet{vis12}: (1) passive (IR) radiative heating of the envelope, (2) heating by UV radiation from the protostar and (3) shock-heating due to the impact of the fast bipolar outflow. They determined that shock heating of envelope cavity walls by the protostellar outflow (to temperatures up to $\gtrsim1000$~K) is the main heating mechanism for far-IR H$_2$O emission. This shock-heating also results in the appearance of highly excited CO lines with $J_u>25$. The fact that our observed CO line fluxes (Table \ref{tab:COfluxes}) fall rapidly with increasing $J$, with no detection of lines with $J_u>24$ indicates a lack of emission from shocked gas along the outflow cavity walls, so strongly sub-thermal shock excitation or radiative heating by the (proto-)star is a likely explanation for the CO line excitation observed in \baboon. The two CO temperature regimes found in \baboon\ are likely indicative of two distinct physical components in the source, perhaps including (but not limited to) a circumstellar disk and warm envelope or an outflow cavity wall component.

The CO rotational diagram is similar to those derived by \citet{kar14}, who obtained best-fitting temperatures $\sim300$~K toward ten high-mass protostars. \citet{kar14} also did not detect the hot, high-$J$ CO component previously observed in low-mass sources, but nevertheless deduced that the observed warm PACS CO emission from massive YSOs probably originates in the heated outflow cavity walls.  

The approximate impact of optical depth on these results may be examined by considering the CO and H$_2$O column densities. Assuming the molecules are evenly distributed across the $\approx10''$ PACS beam surrounding the source (corresponding to a radius of 19,000~AU at 3.8 kpc), the column densities for the warm CO and H$_2$O detected by PACS are $(4.1\pm0.3)\times10^{16}$~cm$^{-2}$ and $(2.4\pm0.3)\times10^{13}$~cm$^{-2}$, respectively. Based on a lower limit to the line FWHM of 3~\kms, derived from a Gaussian fit to the HIFI C$^{18}$O $6-5$ line, the CO lines observed with PACS are all found to have optical depths $\tau<0.5$, and those with $J_u>17$ have $\tau<0.1$. The low-temperature component traced by the lower-$J$ lines may therefore suffer from some self-absorption, whereas the higher-temperature component is likely optically thin. Several of the H$_2$O line fluxes, on the other hand, may suffer more severely from opacity (due to the relatively lower $E_u$ values, which more closely match the apparent excitation temperature of the gas).  The $2_{21}-1_{10}$ and $2_{12}-1_{01}$ lines, for example (at $E_u=194$~K), have $\tau\approx0.7$, and most of the other lines have $\tau\gtrsim0.1$. Self-absorption may thus at least partly explain the scatter observed in the H$_2$O rotational diagram. In addition to optical depth effects, H$_2$O excitation modeling by \citet{vis12} showed that the PACS H$_2$O rotational excitation by outflow shock-heating results in non-LTE level populations and consequent scatter in the rotational diagram.

\subsection{O\,{\sc i} lines}

Integrated fluxes for the O\,{\sc i} $^3$P$_1-^3$P$_2$ (63~$\mu$m) and $^3$P$_0-^3$P$_1$ (146~$\mu$m) lines toward \baboon\ are $82\pm7$~Jy\,GHz and $52\pm2$~Jy\,GHz, respectively, leading to a 63/146~$\mu$m flux ratio of about 1.6. Emission from far-IR O\,{\sc i} lines occurs in hot gas ($T\gtrsim100$~K), where the fine-structure levels are excited by collisions with electrons and atomic/molecular hydrogen, and is typically seen in PDRs \citep[see, for example][]{hol97} and YSOs \citep{lis06,vank10,goi12}, but may also be present in UV-irradiated outflows from evolved stars \citep{gro11}.  The low observed 63/146~$\mu$m flux ratio is difficult to reproduce using single-component excitation models \citep{lis06}, and is at variance with the values of $\sim20$ typically observed in YSOs using {\it Herschel} \citep[\eg][]{wam10}. Self-absorption of the 63~$\mu$m line in tenuous foreground gas was hypothesized by \citet{lis06} as a possible explanation for such anomalously low ratios, which seems plausible given the likely presence of diffuse/translucent gas along the line of sight to \baboon, which passes through the Galactic plane (see Section \ref{sec:higal}).

\subsection{PACS Maps}

\begin{figure*}
\centering
\includegraphics[width=0.49\textwidth]{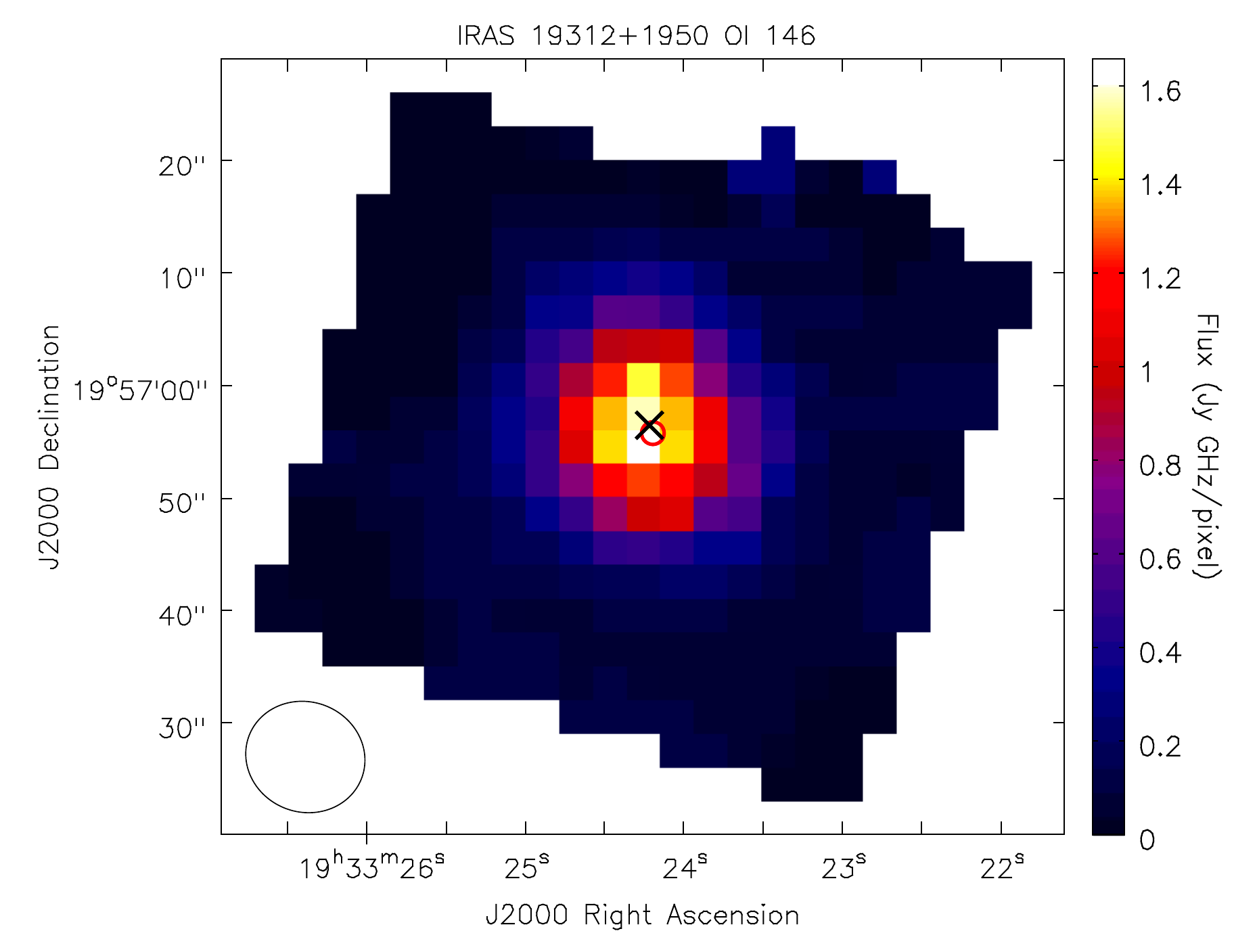}
\includegraphics[width=0.49\textwidth]{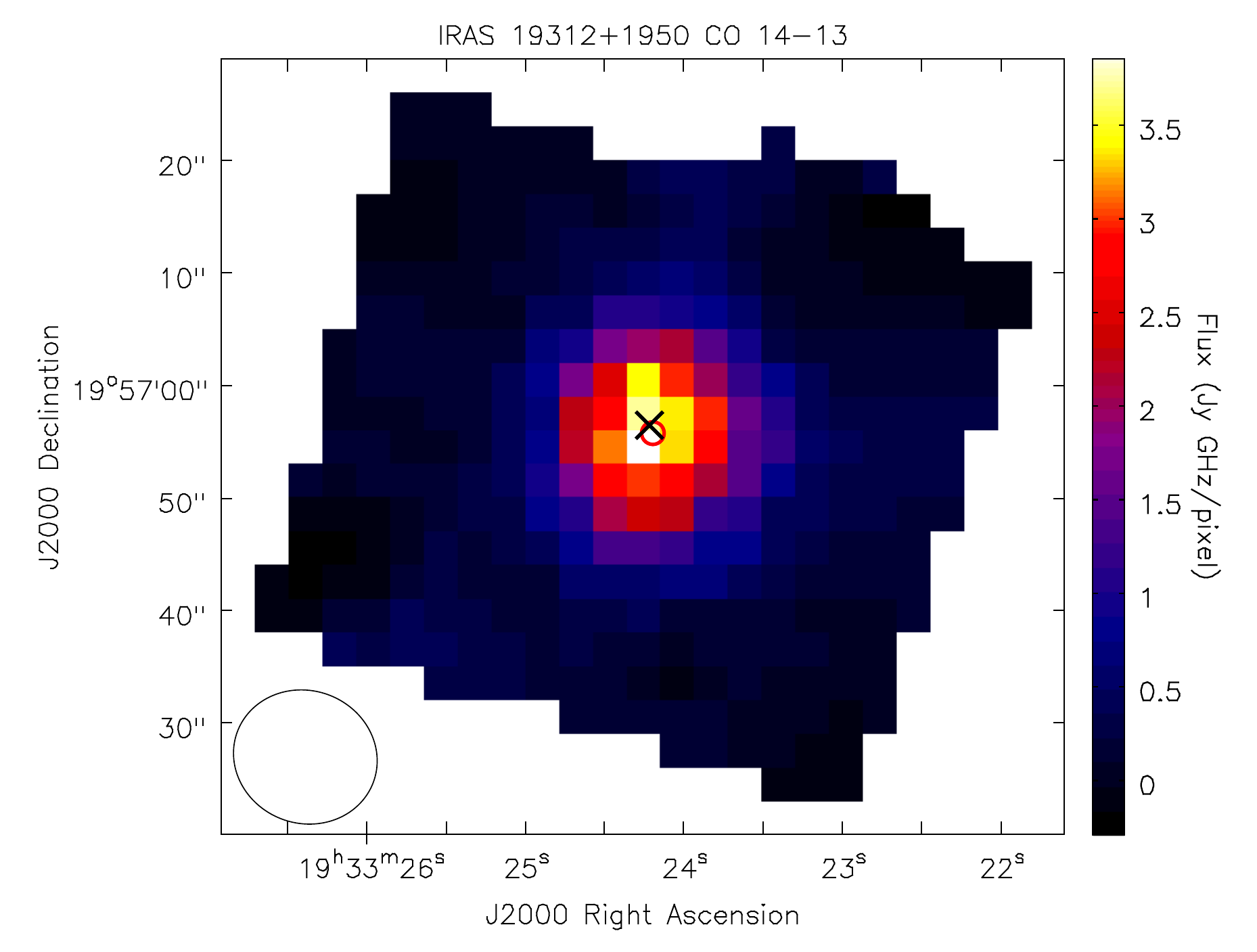}
\includegraphics[width=0.49\textwidth]{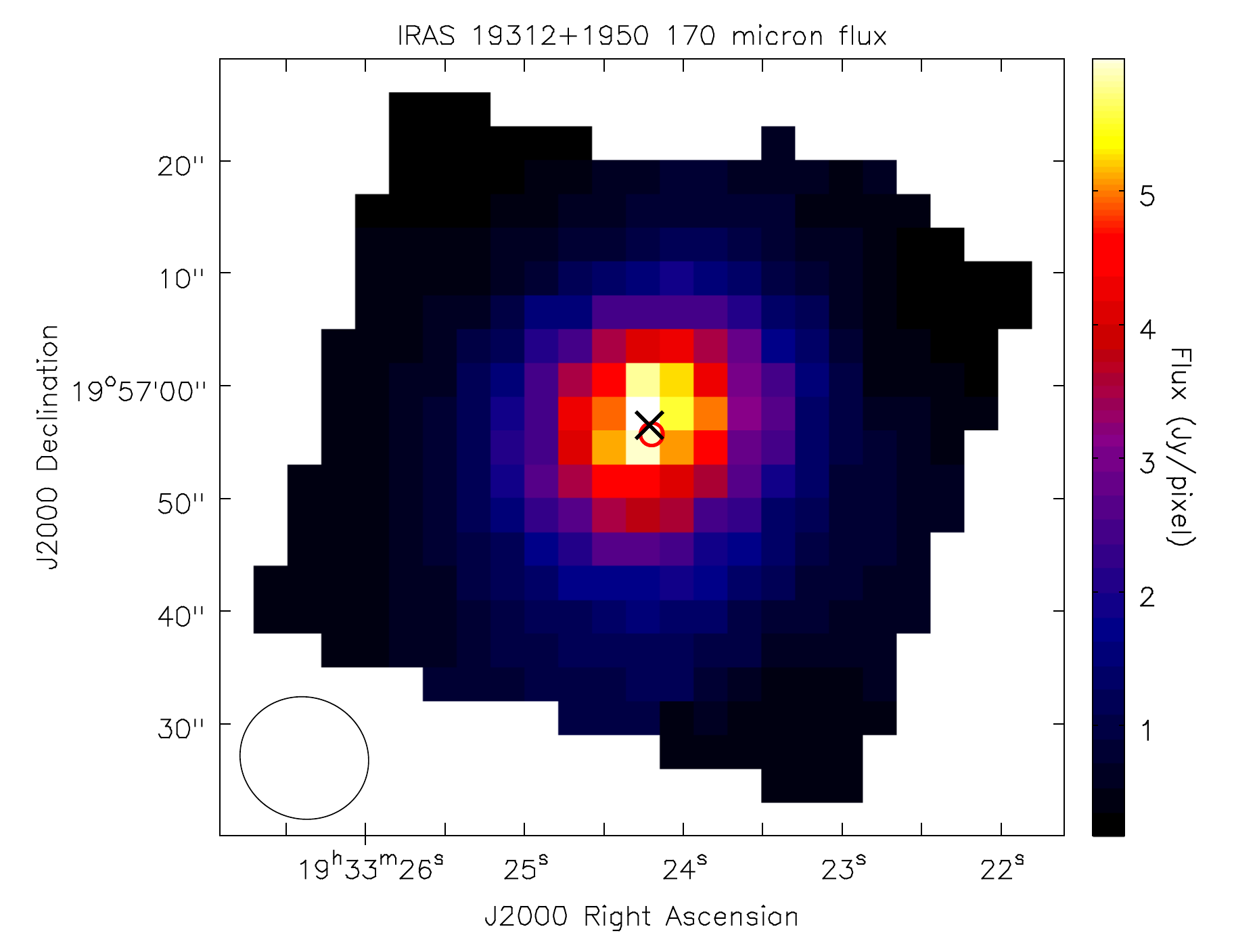}
\includegraphics[width=0.49\textwidth]{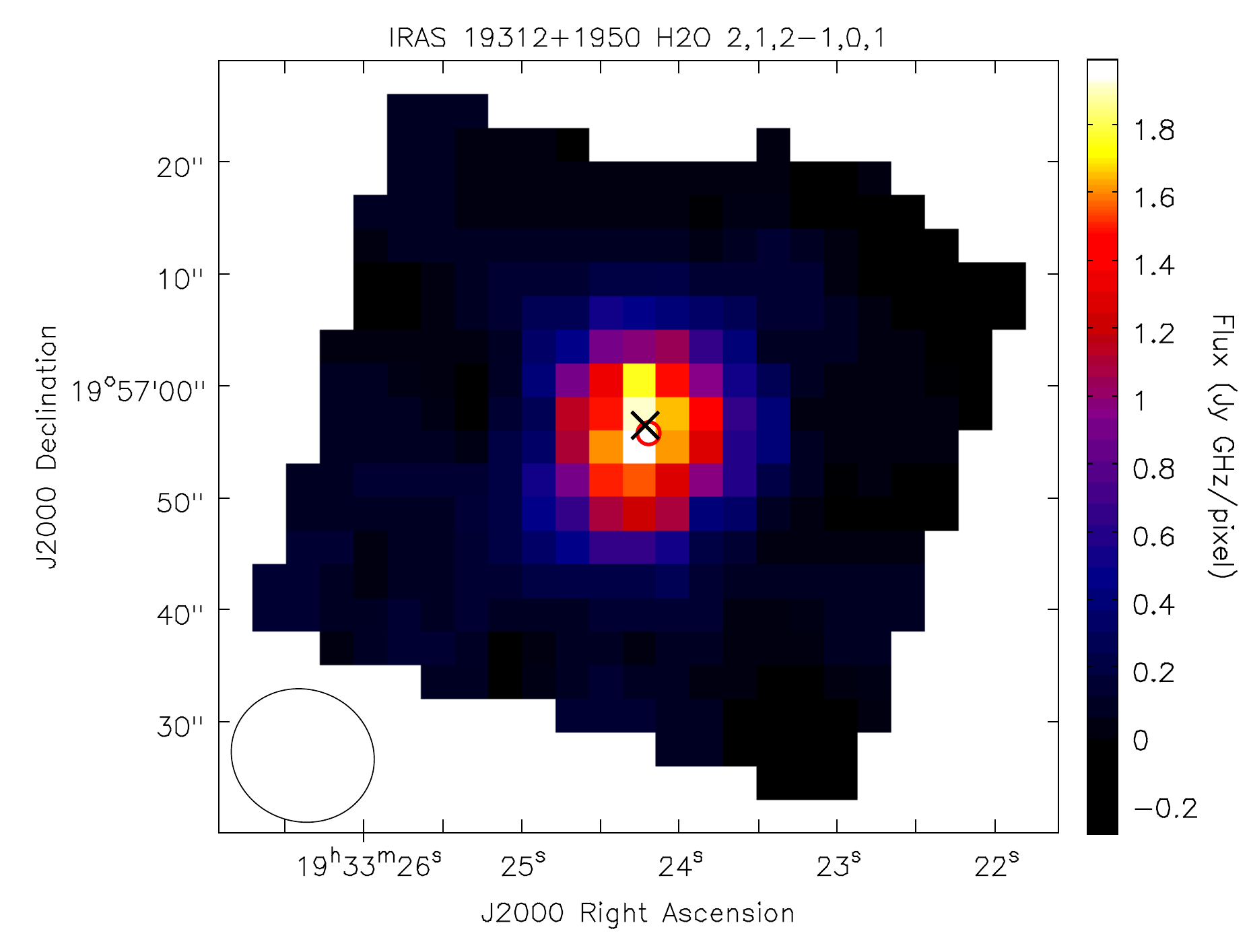}
\caption{Top left: PACS O\,{\sc i} 146~$\mu$m flux raster map. Bottom left: PACS 170~$\mu$m continuum flux raster map. Top right: PACS CO $14-13$ flux raster map. Bottom right: PACS H$_2$O $2_{1,2}-1_{0,1}$ flux raster map. All line flux maps are continuum-subtracted. The FWHM of elliptical Gaussian fits to the PACS PSF are shown lower left in each panel. Black crosses show the position of the peak O\,{\sc i} and 170~$\mu$\,m continuum flux. Red circles indicate the CO (and H$_2$O) far-IR emission peak position. Peak positions were identified using 2-D Gaussian fits.\label{fig:maps} }
\end{figure*}

Spectrally-integrated PACS raster maps for several emission lines of interest, plus the 170~$\mu$m continuum emission, are shown in Figure \ref{fig:maps}. These maps exhibit a striking similarity, consistent with the presence of a dense, compact, dust and gas-rich envelope.

The total emitted flux at at 170~$\mu$m is strong (330~Jy), and peaks at a position consistent with the near-IR source (Fig. \ref{fig:ukidss}). This wavelength traces predominantly cold ($T\sim10$-20~K) dust, and the close spatial correspondence indicates that the near-IR source is likely embedded near the center of a nearly spherical dusty envelope. The 170~$\mu$m morphology is not well resolved but shows evidence for a slightly asymmetric, spatially extended structure, with a long axis FWHM of $23''$, aligned $35^{\circ}$ clockwise from north. The FWHM of the short axis is $20''$, which is still significantly extended compared with the $11''$ PACS PSF. 

Examples of integrated CO and H$_2$O line intensity maps are also presented in Figure \ref{fig:maps}, showing closely similar morphologies between these species, with emission strongly concentrated close to the central source.  These maps also show evidence for elongation in approximately a N-S direction, indicating the presence of an extended molecular envelope or perhaps an unresolved bipolar outflow structure.

Similar to the molecular gas and dust, the O\,{\sc i} 146~$\mu$m emission also peaks up strongly on the central position of \baboon.  Emission from C\,{\sc ii} at 158~$\mu$m was detected in our PACS field in the vicinity of \baboon, with a relatively smooth distribution. However, a similar level of uniform C\,{\sc ii} emission was also detected in the off-source PACS nodding positions $6'$ away. Foreground/background Galactic C\,{\sc ii} emission from the diffuse ISM is a likely explanation for this emission, precluding the use of C\,{\sc ii} as a reliable tracer for ionized gas associated with \baboon.

\subsection{Hi-Gal imaging}
\label{sec:higal}

\begin{figure*}
\centering
\includegraphics[width=0.8\textwidth]{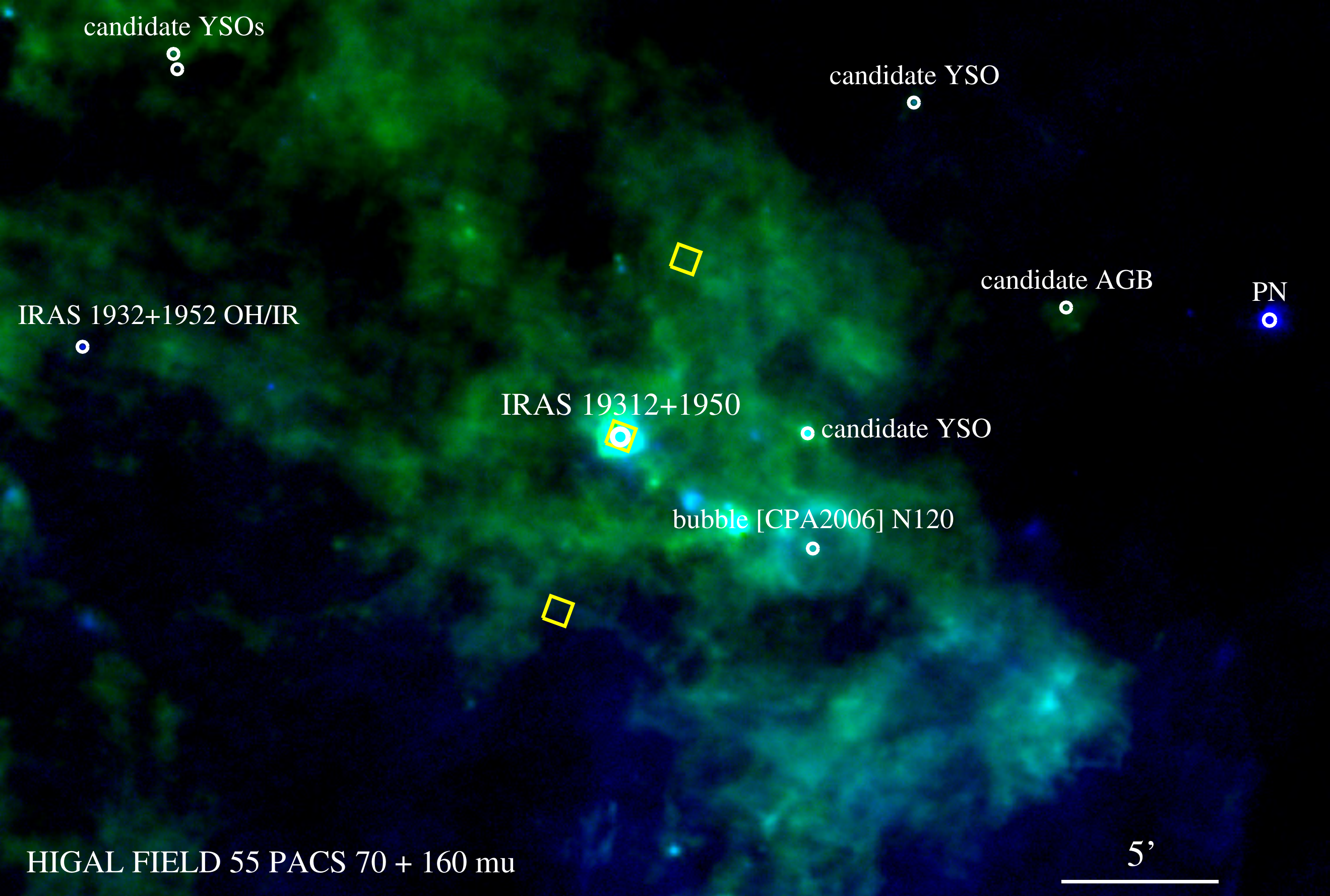}
\caption{{\it Herschel} PACS two-color (green = 70~$\mu$m, blue = 160~$\mu$m) image of the region surrounding IRAS~19312+1950. Previously-identified objects have been labeled. The on-source pointing position and PACS chop/nod reference positions are shown with yellow boxes.\label{fig:higal}}
\end{figure*}

Figure \ref{fig:higal} shows a composite image of PACS 70~$\mu$m and 160~$\mu$m emission in the region surrounding \baboon. This field was extracted from {\it Herschel} observations (OBSIDs 1342219812 and 1342219813), obtained as part of the Hi-Gal Galactic Plane Survey \citep{mol10,tra11}. The maps were reduced using the {\tt Scanamorphos} software \citep{rou13}. Several evolved stars (AGB/PN) and (candidate) YSOs are located within the mapped region \citep{rob08}, indicating past and present star formation in field. Numerous other compact sources nearby await classification. 

Emission in the 70~$\mu$m and 160~$\mu$m wave bands traces predominantly cool and cold dust, and the complex structure across this region highlights the presence of numerous interstellar clouds, clumps, filaments and wind-blown bubbles at various distances along this line of sight through the Galactic mid-plane (at Galactic coordinates $l=55.4^{\circ}$, $b=+0.2^{\circ}$). Despite its large distance ($\approx3.8$~kpc), the Hi-Gal imagery (in addition to 2MASS and UKIDSS) shows that \baboon\ is by far the brightest far-IR source in the region, host to a compact, luminous radiation source surrounded by a mass of cold dust that extends to a distance of at least $30''$ ($\gtrsim10^5$ AU; see also Section \ref{sec:sed}).  

\subsection{HIFI Spectra}

\subsubsection{HIFI Line Fluxes}

Baseline-subtracted, main beam efficiency-corrected HIFI WBS spectra are shown in Figure \ref{fig:hifilines}. Total integrated line fluxes are given in Table \ref{tab:HIFI}.

Due to the low intrinsic abundances of the minor isotopologues $^{13}$CO and C$^{18}$O, the emission from these species should be quite optically thin. Using the ratio of intensities of the $J=10-9$ and $6-5$ lines (employing a two-point rotational diagram, based on the method of \citealt{cum86}), the excitation temperatures and column densities of these species have been derived. For $^{13}$CO and C$^{18}$O, temperatures are found to be $81\pm3$~K, and $66\pm5$~K, respectively, confirming the presence of a gas component cooler than the $\sim200$~K component probed using PACS. Differing temperatures for these two CO isotopologues could be the result of optical depth effects in $^{13}$CO (due to its larger abundance), which would act to reduce the $J=6-5$ line strength relative to $J=10-9$. A differing $^{13}$CO/C$^{18}$O abundance ratio across varying excitation conditions in the source is also possible, for example, if the $^{13}$CO/C$^{18}$O ratio is larger in a hotter region, which could result from isotopic variability in the wind of an evolved star.

Unfortunately, due to the relatively large HIFI beam size (see Table \ref{tab:HIFI}) and the complex source structure containing multiple kinematic components at differing temperatures, little can be inferred about the intrinsic properties of the source from beam-averaged, integrated spectral line fluxes.  Regardless of the nature of \baboon, the presence of excited CO and H$_2$O indicates an intense source of energy. Emission from the higher-energy molecular transitions likely originates in a relatively compact, hot region close to the source whereas lower-energy transitions trace cooler matter further away. Given the uncertain nature of the source, the detailed radiative transfer modeling required for the interpretation of the HIFI line fluxes is beyond the scope of the present article.

\subsubsection{HIFI Line Profiles}
\label{sec:hifi}

\begin{figure}
\centering
\includegraphics[width=\columnwidth]{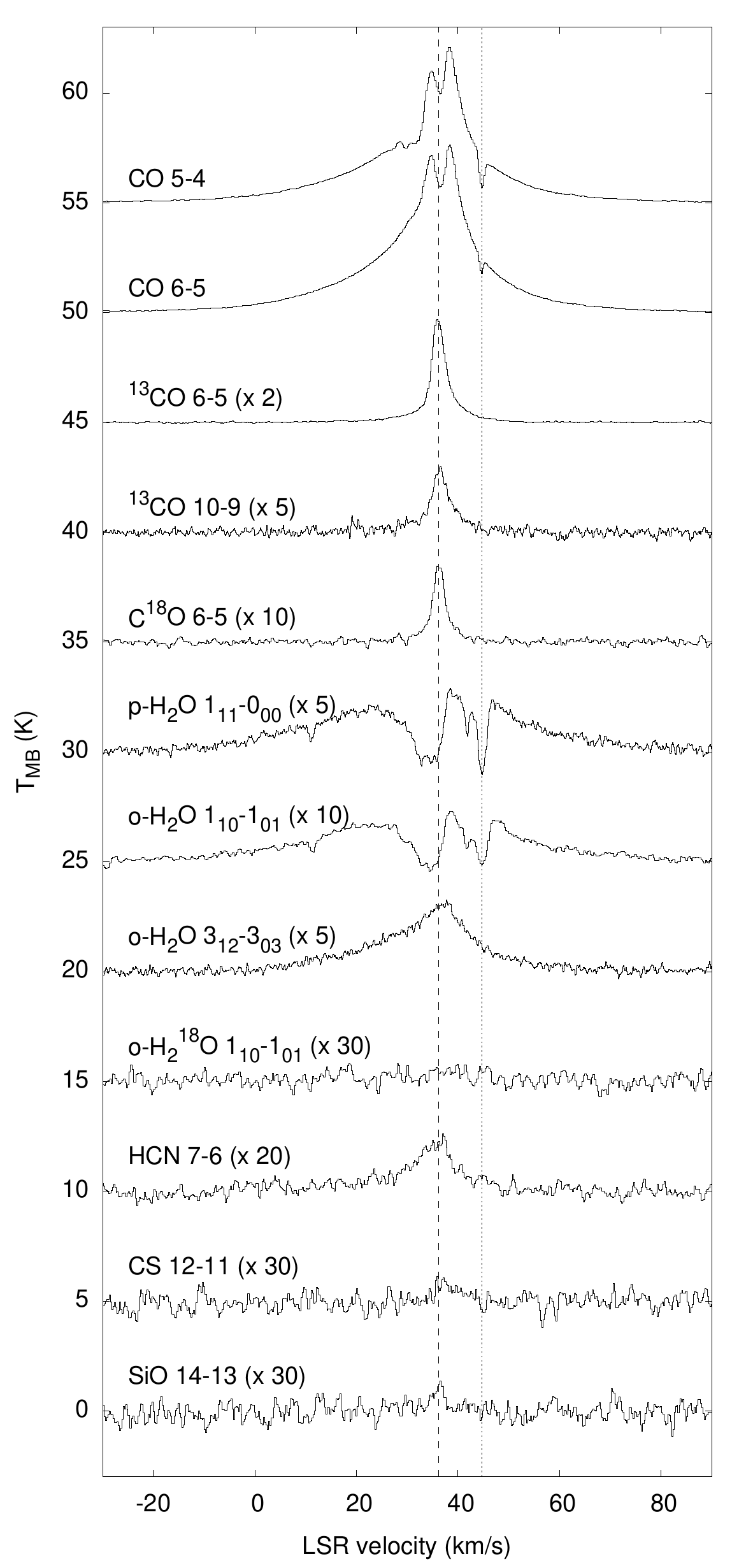}
\caption{HIFI WBS molecular line observations of \baboon. Spectra are baseline-subtracted and have multiplicative scaling (in parentheses), and additive offsets for display.  Dashed vertical line shows the C$^{18}$O $J=6-5$ peak velocity of 36.2~\kms. Dotted line indicates a foreground absorption component. \label{fig:hifilines}}
\end{figure}

The CO and H$_2$O lines in Figure \ref{fig:hifilines} (observed using HIFI WBS), show a wealth of spectral structure, particularly in the lower-$J$ transitions, which trace cooler, more optically thick gas. The CO $J=6-5$ and $J=5-4$ lines exhibit a broad component that rises to a narrow, double-peaked structure near the systemic velocity, indicative of self-absorption in the line cores. The likelihood of CO self absorption is confirmed by the fact that the central dip coincides with the peak velocities of the less optically thick C$^{18}$O and $^{13}$CO lines (at $v_{LSR} = 36.2$~\kms, shown with a dashed vertical line in Figure \ref{fig:hifilines}).

The low energy ($J''=1$) transitions of H$_2$O also exhibit a very broad component which, ignoring the presence of the narrow absorption features, has a relatively smooth and flat-topped, Gaussian-like profile. Meanwhile, the less optically thick H$_2$O $3_{12}-3_{03}$ line has an overall line shape very similar to CO $J=6-5$, for which the broad wings blend smoothly into the narrow line core. Such a `continuous' line profile may be indicative of a continuum of dynamical conditions surrounding \baboon, in which a high velocity outflow mixes gradually into a more quiescent (slowly moving or static) envelope.

The observed HIFI WBS spectra benefit from extremely clean, flat baselines, and emission is detectable far from the line cores: CO $J=6-5$ flux is detected over the range $-65$ to 120~\kms\ (full width at zero intensity, FWZI = 185~\kms) and H$_2$O $1_{10}-0_{01}$ spans $-30$ to 105~\kms\ (FWZI~=~135~\kms). Both ranges are approximately symmetrical about the 36~\kms\ systemic velocity, although the broad, Gaussian-like components of H$_2$O and CO are centered at around 33~\kms\ (blueshifted from the C$^{18}$O peak), indicating asymmetry in the outflow. 

As pointed out by \citet{deg04}, the broad line components imply the presence of a powerful wind emanating from the central star. However, the smooth, broad line profiles observed by HIFI lack the characteristic double-peaked/flat-topped shape associated with (spherical) AGB-star outflows \citep[see for example][]{deb12,kho14,dan14}. Instead their profiles are more similar to those observed towards low and high-mass protostars, where broad line wings are an observational characteristic of shocked gas accelerated by a fast bipolar outflow \citep{kri10,kri12,san16}. Indeed, our observed H$_2$O $3_{12}-3_{03}$ line has FWZI~=~64~\kms, which is consistent with the mean value of $71\pm35$ observed by \citet{san16} in a sample of 19 high-mass YSOs using HIFI. Evolved stars, by contrast, typically have smaller H$_2$O line FWZI in the range 10-45~\kms\ \citep[\emph{e.g.}][]{hun07}.

An approximate temperature for the shocked/outflowing CO can be obtained from the ratio of the $J=6-5$ \emph{vs.} $5-4$ line wing intensities. Excluding the optically thick line core region between 19-48~\kms, a flux ratio of $1.30$ was obtained, which corresponds to an excitation temperature of $\sim330$~K for the accelerated gas. This is roughly consistent with the $\sim300$~K rotational temperature of CO commonly observed by {\it Herschel} in shocked protostellar outflow cavity walls \citep{mot14,kar14,san16}.

A CO absorption component occurs at 44.7~\kms\ (with FWHM~=~0.4~\kms), indicated with a dotted vertical line in Figure \ref{fig:hifilines}, and possibly due to foreground interstellar gas.  This foreground component shows up as a strong absorption feature in the ground-state (ortho and para) H$_2$O transitions, but not in the higher-energy $3_{12}-3_{02}$ line, indicating a relatively low temperature ($\lesssim100$~K). Similar to CO, the ground state ($J''=1$) H$_2$O lines show strong absorption features at around the systemic velocity, but with peak self-absorption offset to the blue, consistent with emission from a warm, inner region and the presence of cooler, outflowing (blueshifted) gas in our line of sight. 

Despite a comparable peak line intensity and similar upper-state excitation energy, the HCN $7-6$ line is significantly broader than the C$^{18}$O $6-5$ line (FWHM~$=13$~\kms\ \emph{vs.} 3.0~\kms), indicating a larger abundance of HCN than CO in the accelerated gas. This indicates more efficient production of HCN than CO in the outflowing/shocked material.

The higher-resolution HIFI HRS spectra are nearly identical to the WBS spectra; the only significant difference is a slight increase in the depth of the narrow absorption component at 44.7~\kms. The HRS spectra are not shown in the present article but are available for download from the {\it Herschel} Science Archive\footnote{Herschel.esac.esa.int/Science\_Archive.shtml}.
                                                                                                                                                                                                             
\subsection{{\it Spitzer} Spectra}
\label{sec:spitz}

\begin{figure}
\centering
\includegraphics[width=\columnwidth]{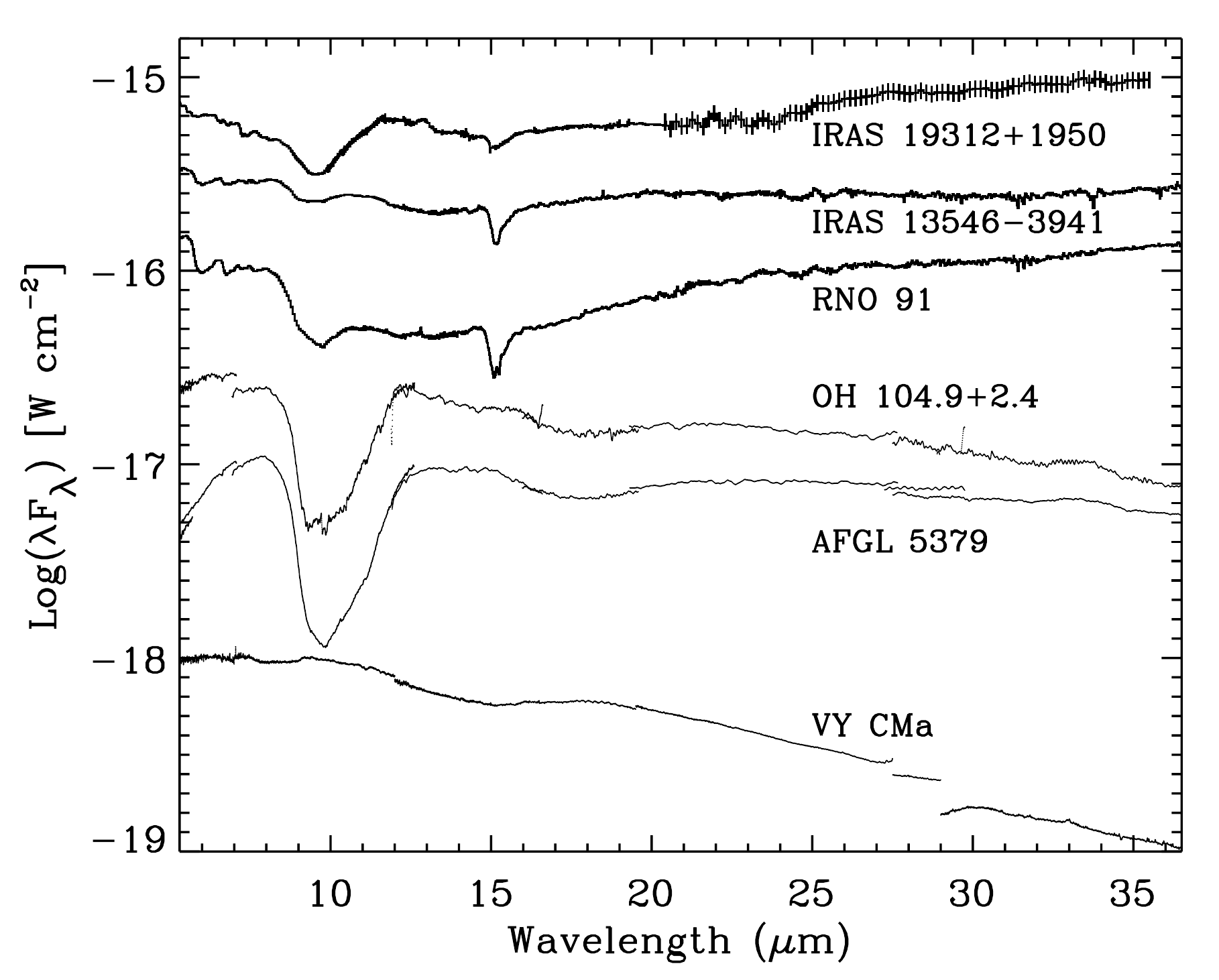}
\includegraphics[width=\columnwidth]{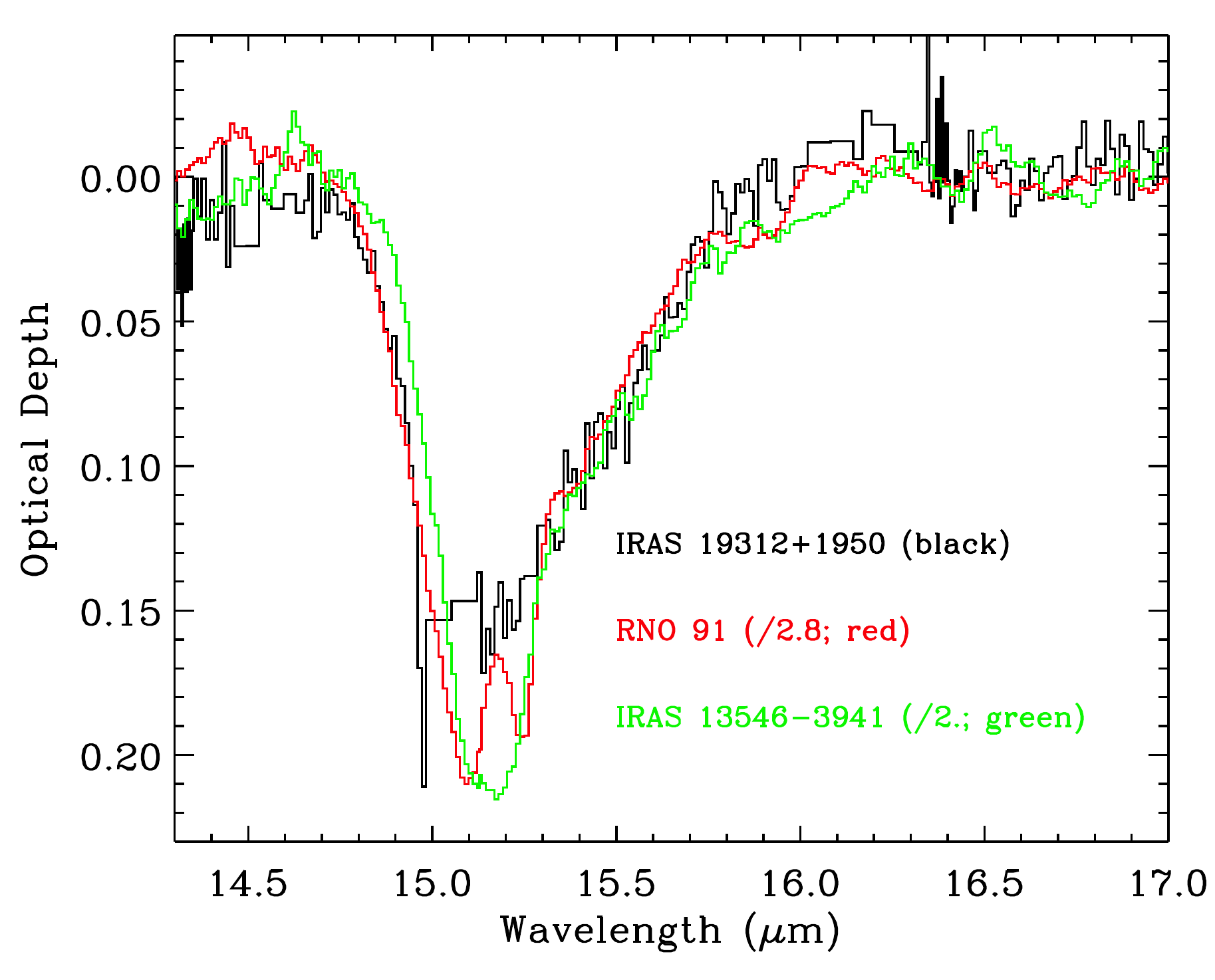}
\caption{Top: {\it Spitzer} IRS spectrum of \baboon\ (top trace), compared with spectra of other YSOs (IRAS~$13546-3941$ and RNO\,91) and evolved stars (OH\,104.9+2.4, AFGL\,5379 and VY CMa) from \citet{boo08}, \citet{syl99} and \citet{har01}. $1\sigma$ error bars are shown for \baboon, which only become significant at wavelengths $>20$~$\mu$m due to detector saturation. Bottom: Zoomed region surrounding the 15~$\mu$m CO$_2$ ice band of \baboon, compared with scaled spectra of other YSOs (overlaid in red and green).\label{fig:spitzer}}
\end{figure}

The {\it Spitzer} IRS spectrum of \baboon\ (covering the range 5-35~$\mu$m) is shown in the top panel of
Fig. \ref{fig:spitzer}. The strongest features, at about 50\% and 15\%, relative to the
continuum, are seen near 9.7~$\mu$m and 15~$\mu$m and correspond to absorption
by amorphous silicates and to the bending vibrational mode of CO$_2$ ice, respectively.
The silicate band may well be affected by emission at the longer
wavelengths, as can be seen by comparison with the spectrum of the low-mass YSO
RNO~91 between 10-12~$\mu$m.

Other, weaker features visible in the {\it Spitzer} IRS spectrum are an
emission band near 12.9~$\mu$m, an absorption near 5.8~$\mu$m, and structure
in the 6.8-7.8~$\mu$m wavelength range, all at around the $\approx7$\% level relative to
the continuum. The 5.8~$\mu$m absorption feature is attributed to the
bending mode of H$_2$O ice. This band is commonly observed at 6.0~$\mu$m toward
embedded YSOs, and the shift to shorter wavelengths indicates that
much of the absorption on the long-wavelength side of this ice feature is filled in
with emission, for example, possibly due to the C-C stretching mode by polycyclic aromatic hydrocarbons. The structure between 6.8-7.8~$\mu$m is most likely real, but at the relatively
low resolution of {\it Spitzer} IRS, it is unclear what the continuum level is, and
thus whether the structure is caused by unresolved emission lines or
absorption features.  An (unidentified) emission feature appears to be present at 12.9~$\mu$m, although the longer-wavelength portion of this feature has been masked out due to saturation of the detector array. In
general, to further study these weaker features, higher spectral resolution ($R\gtrsim1,000$), unaffected by array saturation are needed. Features longward of 20~$\mu$m are within the noise envelope so probably not real.

Crystalline H$_2$O ice absorption bands have been observed in high
mass-loss, dense, evolved stellar envelopes exhibiting OH maser
emission, and also in the expanding, cool envelopes of slightly more
evolved post-AGB stars \citep{boo15}. On the other hand, CO$_2$
ice is not seen in evolved stellar envelopes. For example, toward the
OH/IR star AFGL~5379 (also shown in the top panel of Fig. \ref{fig:spitzer}), the 3.0~$\mu$m H$_2$O ice band has a peak optical
depth of $\sim0.7$ \citep{syl99}, which translates to a H$_2$O
column density of $1.2\times10^{18}$~cm$^{-2}$ (using an integrated band strength of
$2\times10^{16}$~cm$^{-2}$ and a FWHM of 330~cm$^{-1}$). Toward massive YSOs, the CO$_2$/H$_2$O
column density ratio is typically about 19\%, and the lowest observed value
is 11\% \citep{ger99}. Hence, if evolved stars had ISM-like
ice abundances, a CO$_2$ column density of $\gtrsim1.2\times10^{17}$~cm$^{-2}$ would be
expected for AFGL~5379. Using observed band widths \citep{ger99} and laboratory-measured integrated band strengths \citep{ger95}, this would result in peak optical depths of
0.45 for the 4.2~$\mu$m CO$_2$ stretch mode, and 0.05 for the 15~$\mu$m bending
mode. While this is at the noise level for the observed 15~$\mu$m
spectrum (Fig. \ref{fig:spitzer}), it is a factor of 5 above any structure observed
around 4.2~$\mu$m \citep{syl99}, confirming the CO$_2$-poor
environment of OH/IR shells. A similar situation is also true for the evolved O-rich star OH~104.9+2.4 (also shown in Fig. \ref{fig:spitzer}).  We therefore conclude that the 15~$\mu$m CO$_2$ ice
band observed toward \baboon\ must originate from the
envelope/disk of a YSO or a foreground cloud.

For comparison, we also show an ISO spectrum of the high mass-loss red supergiant star VY CMa in the top panel of Fig. 7 \citep[see also][]{har01}. Silicate emission bands occur around 10 and 18~$\mu$m, and the SED falls steadily towards the far-IR, in contrast to the rising SED of \baboon. A close-up of the 15~$\mu$m region oberved by {\it Spitzer} for \baboon\ and the YSOs RNO~91 and IRAS~13546-3941 is shown in the lower panel of Fig. \ref{fig:spitzer}. The close correspondence between these CO ice band profiles confirms a similarity of the conditions in the envelope of \baboon\ to those around other young stellar objects.

\section{Dust emission and SED modeling}
\label{sec:sed}

\subsection{Observational SED data}

\begin{figure}
\centering
\includegraphics[width=\columnwidth]{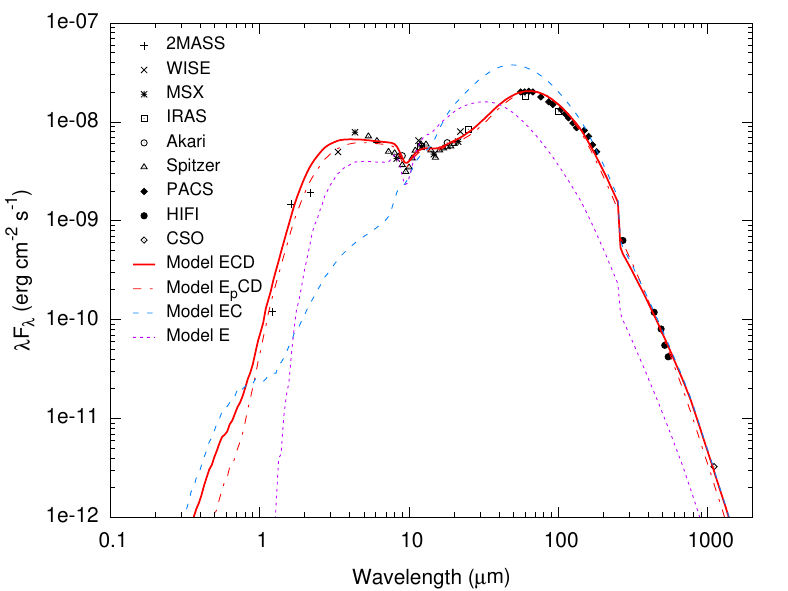}
\caption{IRAS~19312+1950 spectral energy distribution compiled from various telescope observations as shown in the key. Four least-squares YSO SED model fits are overlaid: (1) rotationally-flattened collapsing envelope (E), (2) rotationally-flattened collapsing envelope with bipolar cavity (EC), (3) rotationally-flattened collapsing envelope with bipolar cavity and disk (ECD), and (4) power-law envelope with bipolar cavity and disk (E$_p$CD). Parameters for the best-fitting SED model (using the analytical YSO model of \citealt{whi03}) are given in Table \ref{tab:sed}. Apparent discontinuity in the model SEDs near 250~$\mu$m is due to the difference in aperture size for the PACS and HIFI observations.  \label{fig:SED}}
\end{figure}

Figure \ref{fig:SED} shows a compilation of fluxes measured toward \baboon\ using various (predominantly space-based) telescopes; data from each instrument is plotted using a different symbol. Continuum fluxes were measured from the HIFI spectra, taken from regions adjacent to the spectral lines and free of any apparent line emission in either sideband (see Table \ref{tab:contfluxes}).  The measured HIFI fluxes are average values from two frequencies separated by $\sim10$~GHz due to the contributions from the two receiver sidebands. 

\begin{figure}
\centering
\includegraphics[width=\columnwidth]{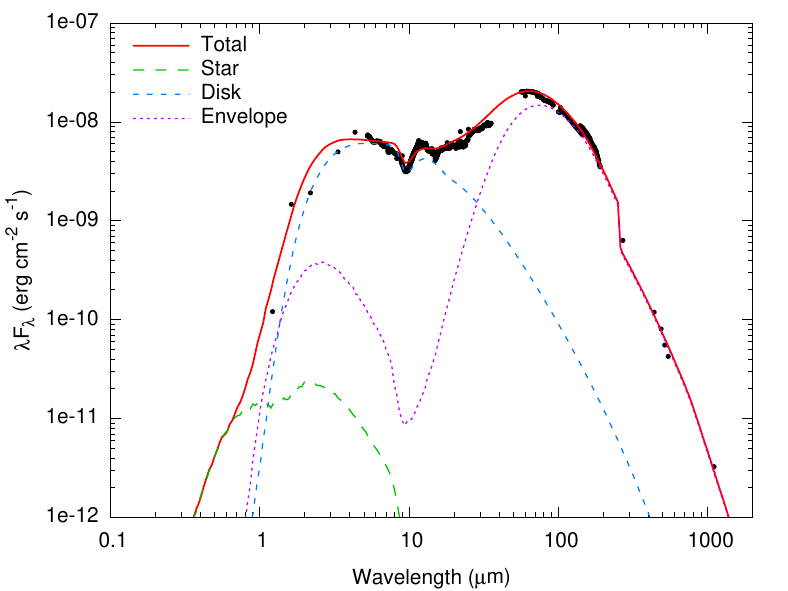}
\caption{IRAS~19312+1950 spectral energy distribution with best-fitting (ECD) model of Figure \ref{fig:SED} overlaid (red solid curve). Contributions to the total SED from the (reddened) central star, the disk and the envelope are shown with dashed curves. Apparent discontinuity in the model SED near 250~$\mu$m is due to the difference in aperture size for the PACS and HIFI observations.  \label{fig:SED_all}}
\end{figure}

\begin{deluxetable}{cl}
\centering
\tabletypesize{\small}
\tablecaption{HIFI continuum fluxes\label{tab:contfluxes}}
\tablewidth{5cm}
\tablehead{$\lambda$ ($\mu$m)&Flux (Jy)}
\startdata
537+549&7.7\\
508+521&9.5\\
484+491&13.1\\
434+440&17.4\\
269+272&57.1
\enddata
\tablecomments{Fluxes are average values for two wavelengths due to contributions from the two HIFI sidebands.}
\end{deluxetable}

The PACS fluxes were extracted from the reduced scan-maps using a $54''$-diameter circular aperture surrounding the central source. The extracted spectra were then uniformly sampled in log-wavelength space, with wavelengths chosen to exclude gaseous emission features and regions of less certain PACS calibration towards the edges of the spectral bands. Despite using the most up-to-date PACS calibration data, slight discontinuities are evident between the different PACS SED spectral bands, although these are within the nominal absolute flux calibration accuracy of 10\%. Reliability of the PACS measurements is confirmed by their close correspondence (within 10\%) of the IRAS measurements at 60 and 100~$\mu$m. The {\it Spitzer} spectra were sampled at a wavelength interval of approximately 0.5~$\mu$m below 11~$\mu$m and 1~$\mu$m above, and fluxes are consistent with previous observations in the mid-IR. Infrared survey fluxes were taken from the most up-to-date data products from each observatory. The Wide-field Infrared Survey Explorer (WISE) 4.6~$\mu$m flux point has been omitted due to detector saturation. The {\it Herschel} PACS and HIFI measurements from the present study lie towards the far right of the plot, sampling the longest wavelengths and filling in the region around the SED peak where data was previously lacking.  The 1,100~$\mu$m data point comes from the Caltech Submillimeter Observatory (CSO) Bolocam survey \citep{ros10}.

The resulting SED (plotted logarithmically in units of uniform energy density, $\lambda{F}_{\lambda}$) shows a broad maximum near 64~$\mu$m, consistent with thermal emission from cold dust.  The 9.7~$\mu$m silicate absorption peak is evident, and a secondary peak around 3-5~$\mu$m is probably due to hot dust in the vicinity of the central source.  Fluxes fall rapidly towards the near-IR, and \baboon\ is extremely faint at optical wavelengths. The double-peaked SED is characteristic of deeply embedded young stellar objects \citep[\eg][]{shu87,rob06,gra09}, whereas evolved star SEDs tend to fall off much more rapidly towards the far-IR \citep[see, for example][]{syl99,har01,pov09,dan14}.

\subsection{SED modeling}
\label{sec:modeling}

Preliminary fitting of the SED was performed using the grid of 200,000 theoretical YSO models published by \citet{rob06}. These SED models consist of four main components: (1) a rotationally-flattened, collapsing circumstellar envelope, (2) a pair of low-density bipolar outflow cavities, (3) a flared Keplerian disk and (4) a central (black body) radiation source. The model parameters span a broad range of stellar temperatures, luminosities, ages, disk, outflow and envelope properties. Additional details of the physical model and its parametrization is given by \citet{whi03}. 

The best-fitting YSO model (\#3017279) was re-computed using the {\sc Hyperion} 3-D Monte Carlo dust radiative transfer code \citep{rob11}. The Kim, Martin \& Hendry (KMH) dust opacity model was used, with a dust sublimation temperature of 1600~K. The number of photons was set to $10^6$ for the specific energy calculation and $10^7$ for peeling-off of the SED fluxes as a function of wavelength. The modified random walk approximation (with $\gamma=1$) was employed for regions of very high optical depth in the disk mid-plane.  The final SED model was calculated using an azimuthally-symmetric polar grid with 200 (altitudinal) angular cells and 500 radial cells, with logarithmic distance increments. 

Due to the decreasing dust temperature with distance from the source, the apparent circumstellar envelope size increases with wavelength, and starts to become significantly larger than the adopted $54''$ PACS aperture for wavelengths $\gtrsim70$~$\mu$m.  For comparison with the PACS data, model SED fluxes were therefore calculated in {\sc Hyperion} using a $54''$ aperture, and for other observations longward of $200$~$\mu$m, an aperture size corresponding to the respective telescope's half-power beam-width (HPBW) was used. For observations at shorter wavelengths, the entire model flux was used (\emph{i.e.}, employing an aperture of infinite size).  

\begin{figure*}
\centering
\includegraphics[width=0.4\textwidth]{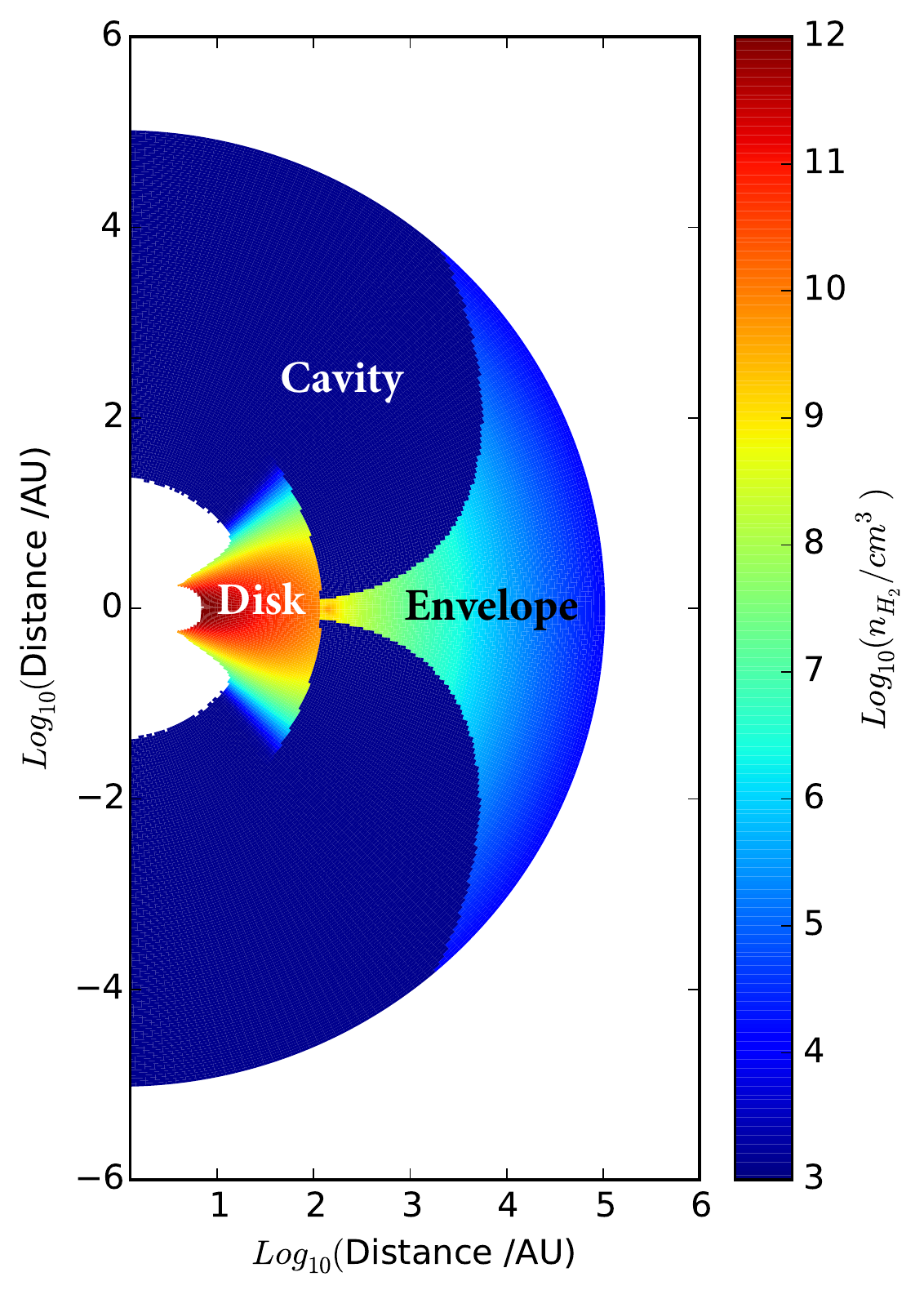}
\hspace{3mm}
\includegraphics[width=0.4\textwidth]{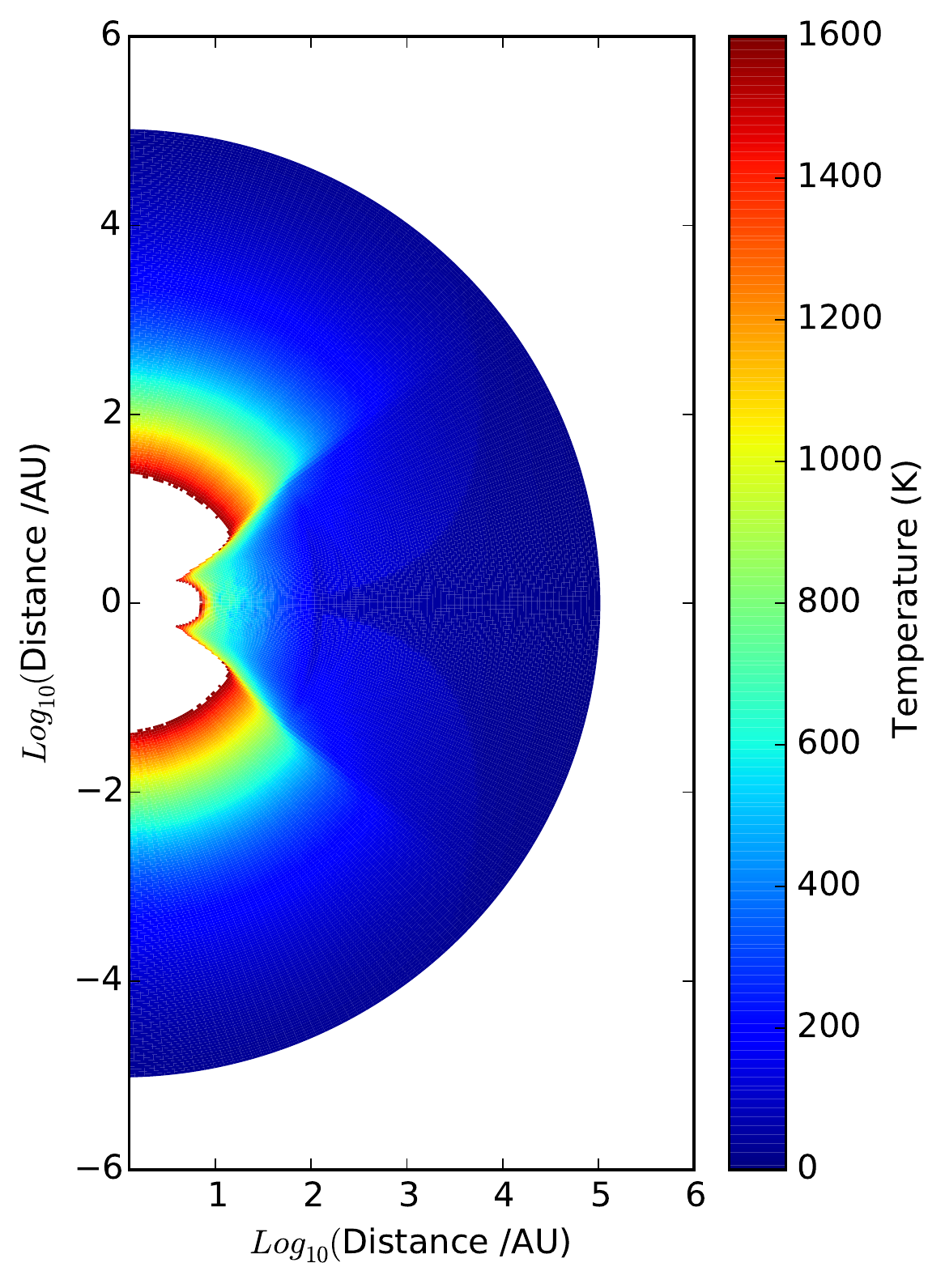}
\caption{Color maps indicating the H$_2$ number density (left) and dust temperature (right) of the best-fitting SED model for \baboon\ (Fig. \ref{fig:SED}), using the analytical YSO model of \citet{whi03} with parameters given in Table \ref{tab:sed}.  The model is azimuthally symmetric. White areas towards the center show where dust has been destroyed due to sublimation at temperatures $>1600$~K. The disk, envelope and outflow cavity are labeled. The distance scale is logarithmic, resulting in apparent distortion of the cavity shape.  \label{fig:sedmodel}}
\end{figure*}

Nonlinear least-squares optimization of the YSO model parameters with respect to the observed SED was performed using the MPFIT routine \citep{mar12}. The distance, inclination angle and foreground extinction were also optimized.  The resulting best-fit SED model is shown in Figure \ref{fig:SED} (solid red curve), and the corresponding parameter values are given in Table \ref{tab:sed}. The model dust density and temperature distributions are shown in Fig. \ref{fig:sedmodel}.

\begin{deluxetable}{ll}
\centering
\tabletypesize{\small}
\tablecaption{Best-fitting SED model parameters \label{tab:sed}}
\tablewidth{0pt}
\tablehead{Parameter & Value}
\startdata
Stellar luminosity & 17,000~$L_{\odot}$\\                                                
Stellar mass       & 9.6~$M_{\odot}$\\                                       
Disk mass          & 0.21~$M_{\odot}$\\                                                       
Disk radius        & 116~AU\\                                                      
Disk scale-height at 100~AU & 17~AU\\                                                       
Disk density exponent & -0.8\\                                                        
Disk flaring power    & 1.1\\                                                        
Envelope infall rate  & $1.23\times10^{-3}$~$M_{\odot}$\,yr$^{-1}$\\                                 
Centrifugal radius & 130~AU\\                                                      
Envelope radius    & $1.1\times10^5$~AU\\                                                 
Cavity opening angle &81$^{\circ}$\\                                  
Inclination angle    & 48$^{\circ}$\\ 
Total mass (gas + dust)&         730$M_{\odot}$\\                                         
Distance           & 3,300~pc\\                                                  
Foreground visual extinction  & 3.7 mag \\
Total visual extinction  & 72 mag                                                  
\enddata
\end{deluxetable}

\subsection{SED model variations}

The requirement of a three-component dust model consisting of an envelope, a bipolar cavity and a disk (abbreviated ECD), was examined by running two additional models: (1) including only an envelope and cavity (EC) and (2) including only the envelope (E). The parameters in these models were subject to the same least-squares optimization as the full ECD model, to arrive at the SEDs plotted using dashed curves in Figure \ref{fig:SED}. The ECD model represents a good fit to the data, whereas the E and EC models are unable to match the observations, thus demonstrating the requirement of (at least) three dust components.  The simplest model (E) is unable to simultaneously reproduce both the mid and far-IR SED peaks because there is no route for radiation to escape and the outer envelope becomes too hot, which shifts the SED peak to the blue. Adding a bipolar outflow cavity with no disk (EC) allows too much radiation to escape from the inner envelope, which is then too cool to reproduce the secondary peak at around 5~$\mu$m. The presence of a dense, inner disk-type structure is required to resolve this discrepancy. To highlight the relative contributions of the three main sources of flux, the separate contributions to the SED from the central star, the disk and the envelope (including both directly emitted and scattered radiation) are shown in Figure \ref{fig:SED_all}.

The final SED model fits the data remarkably well apart from in the mid-IR (between 2 and 4~$\mu$m). Discrepancies in this region could be due to deficiencies in the adopted (azimuthally symmetric) flared disk density distribution. Alternatively, clumpy structure in the outflow cavity, or the presence of a companion star could produce additional sources of opacity or emission in this range.

The hot (up to $\sim1,600$~K), dense (up to $\sim10^{12}$~cm$^{-3}$), dusty disk, residing between about 20 and 150~AU from the central star, provides the main source of flux in our model between 2-20~$\mu$m. However, given the slight mismatch between model and observations at shorter wavelengths, the shape of the disk is certainly not well constrained. Indeed, although a mass of hot, dense dust seems to be required, it may be possible to reproduce this part of the SED without employing a classical disk-like structure. Accordingly, the analytical YSO model (with parameters given in Table \ref{tab:sed}) probably does not represent a unique solution for the structure of the dust distribution. In reality, azimuthal symmetry is unlikely and the structure of the envelope, disk and outflow may be clumpy and turbulent, with various infalling streams and outward-moving jets of material. In addition, the ad-hoc nature of the \citet{whi03} YSO model, creates an artificial (and non-physical) discontinuity at the disk boundaries. Exploring the complete parameter space of physically plausible dust distributions that can reproduce the observed SED may be worthwhile in the future once the region around 1-5~$\mu$m is better characterized, but such an analysis is beyond the scope of the present study. The most important parameters here are the intrinsic luminosity of the central source, as well as the mass and size of the envelope, which are well constrained by our models.  The implied presence of a warm inner disk could be tested through future high resolution sub-mm molecular line observations.

To further assess the uniqueness of the best-fitting SED model parameters, the rotationally-flattened infalling envelope structure of \citet{ulr76}, which has a radial density dependence of $\sim r^{-1.5}$, was replaced by a simple spherically-symmetric envelope with r$^{-\alpha}$ density profile (with $\alpha$ as a free parameter). This model was optimized as in Section \ref{sec:modeling}, and the best-fitting parameters resulted in an envelope mass of $470M_{\odot}$, density exponent $\alpha=2.2$ and source luminosity of $20,000L_{\odot}$. The SED corresponding to this `power-law envelope + disk + cavity' model is shown with a dot-dashed style in Fig. \ref{fig:SED} (and labeled E$_p$CD).  The greater luminosity of this model is primarily due to a slightly larger best-fitting distance of 3.6~kpc. Other parameters such as the disk mass, size and shape and the cavity opening angle and orientation are not significantly different from the best-fitting ECD model parameters in Table \ref{tab:sed}.  Given the known distance of \baboon\ from H$_2$O maser parallax observations ($3.8^{+0.83}_{-0.58}$~kpc; \citealt{ima11}), the total luminosity is well constrained to the range $L=16,000$-$25,000L_{\odot}$. 

A black-body temperature for the central source in the range $T\sim10,000-25,000$~K provides the best fit to the SED. Lower temperatures result in a worse fit in the 20-30$~\mu$m region, but cannot be ruled out given the lack of observational constraints on the structure of the disk and inner envelope. Indeed, if \baboon\ is a YSO then a relatively low source temperature (and young evolutionary stage) is implied by the non-detection of free-free radio emission. Using the formula of \citet{car90}, the $0.66$~mJy $3\sigma$ upper limit of the CORNISH 5~GHz survey \citep{pur13} corresponds to a total ionizing photon flux of $10^{45}$\,s$^{-1}$, which suggests a relatively small H\,{\sc ii} region with a stellar effective temperature $\lesssim10,000$~K.

Our best-fitting SED model differs significantly from that of the AGB-envelope model of \citet{mur07}, who derived $L=7,000L_{\odot}$ for an adopted source distance of 2.5~kpc. The discrepancy between our models is primarily due to improvements in the characterization of the far-infrared fluxes by {\it Spitzer} and {\it Herschel}, but also to the improved distance estimate from \citet{ima11}.  Although \citet{mur07} took the IRAS far-IR fluxes at 60 and 100~$\mu$m into account in their modeling, they assumed a large background contamination (and hence, a large uncertainty) for those measurements. Our {\it Herschel} images demonstrate that the assumption of a strong far-IR background was erroneous, as shown in Figures \ref{fig:maps} and \ref{fig:higal} where \baboon\ appears as a spatially isolated object. Taking the average background flux from the Hi-Gal 70~$\mu$m image using an annulus between 150-$200''$ from the center of the source gives a possible background flux contribution of only 3.4\% for our measured far-IR continuum fluxes. Furthermore, as shown in Figure \ref{fig:higal}, the PACS reference positions (yellow squares) covered regions with a similar level of background emission to that near the central source, thus providing a (serendipitous), approximate background correction for the source signal and negating the need for additional corrections.

\section{Discussion}

The circumstellar envelope mass of $\sim500$-$700M_{\odot}$ derived by modeling the far-IR SED compares reasonably well with the 225-478$M_{\odot}$ obtained from CO observations by \citet{nak16}, and is much too large to have originated from the outflow of an evolved star. \citet{nak16} derived their envelope mass assuming a standard interstellar gas-phase CO/H$_2$ ratio of $8\times10^{-5}$, so their total mass estimate would be correspondingly larger if a fraction of the CO is frozen out onto dust (which is usually the case in massive star-forming cores; \citealt{ryg13}).  \citet{deg04}'s original envelope mass estimate of only $25M_{\odot}$ (from CO $J=1-0$ line modeling, assuming a $r^{-3}$ density distribution between $r = 5$-$15''$) appears to have severely underestimated the amount of gas surrounding \baboon. A clump mass of $1200\pm200M_{\odot}$ was obtained by \citet{dun11} using 1.1~mm continuum and NH$_3$ radio spectroscopic observations of \baboon\ (assuming a spherical source with constant temperature and density), which provides further evidence for a large total mass of circumstellar material. In our best-fitting SED model, circumstellar material contributes $\sim70$ mag. towards the total line-of-sight visual extinction of the central star. Such high extinction implies that the majority of the observed CO$_2$ and H$_2$O ice absorption occurs in the circumstellar envelope rather than in foreground interstellar material.

The stellar luminosity and mass derived as independent parameters from our SED fitting are consistent with the pre-main-sequence (PMS) evolutionary track of an early B-type supergiant star \citep[\eg][]{her04,eli10}. Based on massive YSO accretion models, the large luminosity and intermediate envelope mass would place \baboon\ close to (within a few times $10^4$ years of) the end of its main accretion phase \citep[see for example][]{mol08}. Such high luminosities are rarely observed in massive evolved stars due to the very rapid evolution during the final stages of their lifetimes (see Figure 2 of \citealt{van03}); a $7M_{\odot}$ post-AGB star is predicted to expel its envelope in only a few hundred years, although thermal pulses can prolong this stage.

The average CO temperature observed with PACS of $T=197.7\pm2.1$~K, and outflow temperature from HIFI of $\sim330$~K, are consistent with the $\sim300$~K temperatures observed in the the outflow cavity walls of low and high-mass protostars derived in other {\it Herschel} studies \citep[\eg][]{mot14,kar14,san16}. The broad, Gaussian-type line wings observed in our H$_2$O and CO HIFI spectra are also highly characteristic of shocked outflowing protostellar gas. \citet{san16} found that the majority of H$_2$O $3_{12}-3_{03}$ line emission from massive YSOs originates in shock-accelerated gas along the cavity walls. In addition to the close similarity in the H$_2$O $3_{12}-3_{03}$ line profile between \baboon\ and other protostars, the integrated line luminosity of $L_{\rm H_2O}=2.0$~K\,\kms\,pc$^2$ (combined with a source luminosity of $\sim20,000L_{\odot}$), puts \baboon\ precisely in line with the $L_{\rm H_2O}$ \emph{vs.} $L_{bol}$ trend observed in 32 protostars by \citet{san16}. Similarly, the CO $J=14-13$ and O\,{\sc i} 146~$\mu$m line luminosities (0.60 and $0.23L_{\odot}$, respectively) are consistent with the close correlations between line luminosity and stellar luminosity observed by \citet{kar14} in a large sample of protostars spanning the range from low to high mass. Thus, the same conclusion regarding the nature of \baboon\ is reached through three separate lines of analysis: (1) the low-resolution far-IR spectrum, (2) the high-resolution HIFI line profiles and (3) the infrared SED, which are all consistent with a luminous protostar embedded in a massive duty envelope, and powering a fast bipolar outflow. The power-law envelope density structure that provides a good fit to the SED of \baboon\ is a common feature of actively-accreting protostars \citep{tob15,ket15}. Based on our interpretation of the data, the idea that a massive evolved star found its way to the center of a collapsing molecular cloud during its brief, final, high-luminosity phase, is highly unlikely.

Two temperature regimes are seen in our PACS CO observations.  The warmest component (at 280~K) has a mass of around $0.22M_{\odot}$, which is similar to the mass of the disk required to match the mid-IR continuum emission in our best-fitting SED model. The cooler (160~K) component is associated with $\approx1.6M_{\odot}$ of material, which might be representative of shocked or radiatively heated gas in the outflow cavity walls further from the star. This hypothesis could be examined by mapping the thermal and kinematic structure through spectrally and spatially resolved observations of high-excitation CO lines.  Such observations would be invaluable as a probe of the energetic feedback processes between star and envelope and would facilitate understanding the nature of outflow shocks in massive YSOs.

Our new understanding of \baboon\ is consistent with the previous molecular line observations of this source. The $\sim2\times10^5$~AU ($\sim50''$ diameter) dense, dusty circumstellar envelope lies within the boundaries of a $\sim100''$-wide CO cloud detected at the same position by \citet{nak16}.  Broad ($\sim50$~\kms\ wide) components in the CO $J=1-0$ and $2-1$ lines observed by \citet{nak05} are consistent with the outflow wings we observed using HIFI, although these were previously interpreted as arising in a spherically-expanding (AGB star) outflow. It was noted by \citet{nak05}, however, that their line profiles were rather inconsistent with the flat-topped shape expected for a spherical outflow.  In the collapsing protostellar envelope paradigm, the \citet{nak05} CO observations may be best explained by a wide-angle bipolar outflow. If the outflowing gas decelerated as it expanded into the ambient envelope, this would be consistent with the observation of a spatially compact, high-velocity CO component that gradually transitions to slower, more extended components with distance from the central star. This explanation may be consistent with the smooth transition from high-velocity wind to quiescent (at rest) envelope, manifested in our HIFI $^{12}$CO and $^{13}$CO $J=6-5$ line profiles. The narrow, 1-2 \kms\ CO outflow component identified by \citet{nak05} could plausibly be probing a slower-moving part of the accelerated gas further out in the envelope. This is consistent with the $1-0$ transition probing mostly cooler gas, which would be further from the shock front of the putative protostellar outflow. Alternatively, the low-velocity (narrow) red and blue-shifted CO components could originate in the receding and approaching hemispheres of a rotating, collapsing envelope or massive disk. Other high-velocity (broad) molecular line components observed by \citet{deg04} are readily explainable as arising from shocked gas from a protostellar envelope, swept up by an impinging fast bipolar outflow.

Two red and blue-shifted H$_2$O (and SiO) maser components were observed towards \baboon\ by \citet{nak11}. The H$_2$O maser components are separated by about 11~mas along an axis $107^{\circ}$ counterclockwise from north, and interpreted as most likely arising in a bipolar outflow. Water maser activity is commonly associated with massive star formation \citep{chu90} as well as outflows from evolved stars \citep{yoo14}. As shown by \citet{tor98}, H$_2$O masers can be associated with both protostellar outflows and disks.  If the H$_2$O (and SiO) maser spots of \baboon\ are interpreted as arising in a Keplerian disk with its axis in the plane of the sky, then the velocity displacement of 35~\kms\ corresponds to a mass of 4.4-6.7$M_{\odot}$ for the central star (assuming a distance of 2.5-3.9~kpc; \citealt{nak11}). Accounting for a $48^{\circ}$ inclination to our line of sight (as derived from our SED modeling), yields a stellar mass range 6-9$M_{\odot}$, which is consistent with a massive protostar. Alternatively, due to the strongly linear distribution of H$_2$O maser spots, with a clear separation between the red and blue components, their association with a well-collimated bipolar outflow may be more likely. In this scenario, the apparent N-S extension of our PACS spectral line and continuum images (Fig. \ref{fig:maps}) could be representative of the size of the wide-angle cavity in a direction perpendicular to the outflow and partially oriented towards our line of sight. Additional high-resolution molecular mapping, as well as continued monitoring of the positions of the H$_2$O maser spots is needed in order to establish the direction of the bipolar outflow and disk axes, to help resolve this dichotomy. 

Previous radio studies have shown 18~cm OH maser line properties to be useful diagnostics of YSOs and evolved stars \citep[\eg][]{her85,tel96,cas98,edr07}. Accordingly, the strength of the 1612~MHz OH maser line relative to the 1665 and 1667~MHz lines observed by \citet{nak11} was taken as good evidence that \baboon\ is an evolved star and not a YSO. Although the 1612~MHz OH maser tends to appear most strongly in the expanding shells of high mass-loss oxygen-rich AGB stars \citep{dav93}, it can also be found in massive YSOs \citep[\emph{e.g.}][]{coh06,ram06}. The precise physical conditions required for pumping this transition are not yet fully understood \citep{fis06}, so a closer examination of the cause of its unusually strong occurrence in \baboon\ could help shed light on this maser's origin.

\section{Classification of \baboon\ and comparison with other sources}

Our new infrared observations of \baboon\ have revealed a far-IR spectrum of gas, dust and ice highly consistent with a massive YSO in an early evolutionary stage. The hypothesis that the central radiation source could be an evolved star inside its natal molecular cloud \citep[as suggested][]{nak11,nak16}, is deemed very unlikely. No such objects have been previously identified, which is consistent with the expectation that the protostellar envelope and natal cloud(s) would have dissipated and returned to the diffuse phase during the latter stages of star formation \citep{lad87,mck07}. 

The possibility of a chance encounter between a massive evolved star and a dense molecular cloud cannot be ruled out entirely. However, the red supergiant (RSG) scenario favoured by \citet{nak16} seems unlikely due to the $\gtrsim90$~\kms\ outflow velocity implied by our HIFI CO observations. Terminal RSG wind speeds are typically $\sim10$-30~\kms, and based on the relation of \citet{mau11}, the relatively low luminosity of \baboon\ (compared to other RSGs), would imply a wind speed of only about 11~\kms. More highly-evolved post-AGB stars (such as OH\,231.8+4.2), on the other hand, can drive bipolar outflows with speeds up to several hundred \kms, but as previously mentioned, the duration of this phase is very short for objects more luminous than $\sim10^4L_{\odot}$, and thus rarely seen.

The well-studied DR21(OH) massive star-forming region makes a useful point for comparison as it shares several similarities with \baboon, with a similar bolometric luminosity, molecular emission line spectrum (including total CO and H$_2$O line luminosities) and far-IR SED \citep{kar14,jak07}.  DR21(OH) also exhibits Class~I CH$_3$OH masers and a tentative SiO maser detection \citep{kal10}, but without a strong 1612~MHz OH maser. Similar to \baboon, the lack of H$\alpha$ emission in DR21(OH) \citep{kum07} is consistent with a young evolutionary stage (and low effective temperature), implying that a large H\,{\sc ii} region has not yet developed. 

Orion KL Source~\emph{I} is a nearby massive protostar that possesses SiO, H$_2$O and OH (1612 and 1665~MHz) masers, as well as a warm disk and bipolar outflow \citep{gre04,coh06,pla09}.  The apparent similarity of these characteristics with \baboon\ suggests a similar evolutionary stage. By this analogy, the origin of the SiO and OH 1612~MHz masers in \baboon\ may also be in the (wide-angle) bipolar outflow \citep[\emph{e.g.}][]{coh06}. Further studies of the structure of \baboon\ could be helpful to improve understanding of the kinematically complex region surrounding Source~\emph{I}.

As highlighted by \citet{nak16}, the environment surrounding \baboon\ does not appear to show any signs of prior star formation. The lack of other identified protostars nearby could indicate that this object is a relatively isolated site of high-mass star formation, but more detailed IR imaging studies are required to properly determine the young stellar population in this region.

\section{Concluding remarks}

The observational evidence presented in this article leads to a self-consistent view of \baboon\ as a massive YSO embedded in a collapsing molecular envelope. Far infrared SED modeling and comparison of the IR spectral features with other sources provides strong evidence to support this scenario. Indeed, the spectrum of PACS and HIFI emission lines is very similar to that observed previously in other massive protostars. The fact that SiO and OH maser observations of \baboon\ have previously been considered to be more characteristic of an evolved star highlights the unusual nature of this object, making it an ideal candidate for followup observations (using ALMA, SOFIA, and JWST, for example), to confirm its identity and search for any other peculiarities that may help inform our understanding of the process of high-mass star-formation / stellar evolution.

Future studies to confirm the nature of \baboon\ will require high-resolution imaging to elucidate the spatial and kinematic structure of the outflow, envelope and putative disk. Observations of optically thin, dense molecular gas tracers such as C$^{18}$O, CS, HCN and H$_2$CO using sub-mm interferometry (at sub-arcsecond resolutions) should be particularly revealing. Detection and mapping of outflow tracers such as CO, SiO and HCO$^+$, and `hot core' chemical tracers such as CH$_3$CN, CH$_3$OCHO and C$_2$H$_5$CN would also help confirm the YSO identification. The presence of a compact H\,{\sc ii} region may be revealed by searching for emission from hot, ionized gas, either through deep radio continuum observations or far-IR line searches for C\,{\sc ii}, N\,{\sc ii} and other ions. 

Although our data strongly indicate the presence of a massive YSO, a chance coincidence with a massive evolved star along the line of sight still cannot be ruled out. The presence of an AGB star could be established for example, by measuring the profile of the 3~$\mu$m H$_2$O absorption band to determine the presence of crystalline H$_2$O. More detailed mapping of far-IR emission from dust, O\,{\sc i} and H$_2$O would also be worthwhile to elucidate the energetic environment close to the star.

\acknowledgements

Support for this work was provided by NASA through an award issued by JPL/Caltech and through NASA's Origins of Solar Systems program. We gratefully acknowledge the work of Thomas Robitaille for providing and supporting the Hyperion radiative transfer code.

{\it Facilities:} \facility{Herschel Space Observatory}, \facility{Spitzer Space Telescope}

\bibliographystyle{apj}
\bibliography{refs}   

\clearpage


\end{document}